\documentclass{svjour3}                     
\smartqed  

\usepackage{amsmath}%
\usepackage{amsfonts}%
\usepackage{amssymb}%
\usepackage{graphicx}
\usepackage{color}

\usepackage{subfigure}
\usepackage[colorlinks=true]{hyperref}
\hypersetup{urlcolor=blue, citecolor=red}

\def\RR{\mathbb{R}}
\def\N{\mathbb{N}}
\def\R{\mathbb{R}}
\def\PP{\mathbb{P}}
\def\EE{\mathbb{E}}
\def\E{\mathcal{E}}
\def\B{\mathcal{B}}

\newcommand{\F}{\mathcal{F}}

\newtheorem{theo}{Theorem}

\newtheorem{prop}[theo]{Proposition}

\numberwithin{equation}{section}

\definecolor{orange}{rgb}{1.00,0.50,0.0}	
\definecolor{green}{rgb}{0.0,0.60,0.00}	

\DeclareMathAlphabet{\mathonebb}{U}{bbold}{m}{n}

\begin{document}
\title{Limits and dynamics of randomly connected neuronal networks}
\author{Cristobal Quininao \and Jonathan Touboul}
\institute{
J. Touboul and C. Quininao \at
Mathematical Neuroscience Team, CIRB-Coll\`ege de France\footnote{CNRS UMR 7241, INSERM U1050, UPMC ED 158, MEMOLIFE PSL*}\\
11, place Marcelin Berthelot\\
75005 Paris, FRANCE\\
\email{cristobal.quininao@college-de-france.fr\\}
\email{jonathan.touboul@college-de-france.fr}
\and INRIA BANG Team, Paris-Rocquencourt Center
\and C. Quininao \at
Laboratoire Jacques-Luis Lions, Universit\'e Pierre et Marie Curie\footnote{UPMC Univ Paris 06, UMR 7598, Laboratoire Jacques-Louis Lions, F-75005, Paris, France}\\
4, place Jussieu\\
75005 Paris, FRANCE
}

\date{Received: date / Accepted: date}

\maketitle

\begin{abstract}
Networks of the brain are composed of a very large number of neurons connected through a random graph and interacting after random delays that both depend on the anatomical distance between cells. In order to comprehend the role of these random architectures on the dynamics of such networks, we analyze the mesoscopic and macroscopic limits of networks with random correlated connectivity weights and delays. We address both averaged and quenched limits, and show propagation of chaos and convergence to a complex integral McKean-Vlasov equations with distributed delays. We then instantiate a completely solvable model illustrating the role of such random architectures in the emerging macroscopic activity. We particularly focus on the role of connectivity levels in the emergence of periodic solutions. 
\end{abstract}

\newpage

%

\tableofcontents

\medskip

\hrule

\section*{Introduction}
Neuronal networks in the cortex are composed of large structures, called cortical columns, that are in charge of collective information processing. Neurons are characterized by a nonlinear activity subject to an intense noise. They interact by sending action potentials (spikes) to those neurons they are connected to. The transmission of the information takes a specific time, related to the characteristic time of the synaptic chemical machinery and to the transport of signals at finite speed through the axons (and therefore function of the anatomical distance between the cells). 

The macroscopic behaviors emerging from such large-scale systems provide relevant signals that are recorded by usual imaging techniques and from which physicians can infer hallmarks of function and dysfunction. Large-scale networks are therefore adequate scales to uncover the function of the cells, and as such have attracted much work in the past few years. Indeed, while properties of single cells have been well known since the seminal works of Hodgkin and Huxley~\cite{hodgkin-huxley:39,hodgkin-huxley:52}, models of macroscopic behaviors are less understood and computational studies have mainly relied on heuristic descriptions of macroscopic behaviors through firing-rate models, following the important work of Wilson and Cowan (WC)~\cite{wilson-cowan:72,wilson-cowan:73}. In this class of models, we will make a distinction between macroscopic models in which the activity considered describes a whole brain area (which correspond to finite-dimensional WC systems) and mesoscopic models that describe macroscopic variables at a finer scale at which averaging effects occur but where we can resolve finer structure of the brain (e.g., WC integro-differential neural field equations). WC models have been very successful in reproducing a number of relevant phenomena in the cortex such as visual hallucinations, which was related to symmetry breaking and pattern formation in the neural field equation~\cite{ermentrout-cowan:79} and binocular rivalry in macroscopic models~\cite{shpiro2007dynamical}, see~\cite{bressloff:12} for a recent review. WC model describes the evolution of a macroscopic variable, the population-averaged firing-rate, as a deterministic variable, which satisfies a delayed differential (macroscopic scale) or integro-differential (mesoscopic scale) equation. The success of these models prompted much work in order to provide a link between such macroscopic regimes and the noisy activity of individual cells. Mean-field methods based on the statistical physics theory of gases was also used for biologically relevant neuronal models~\cite{toubouldelays:12,touboulNeuralfields:11} including noisy input or noisy synaptic transmission and delays. Similarly to the molecular description of gases, it is shown that the propagation of chaos property takes place and that the system converges to a particular class of McKean-Vlasov equations.

In the vast majority of these studies, networks are assumed to be fully connected (i.e. all cells interact together), and no specific topology is taken into account. If this assumption is relevant in the molecular theory of gases, such architectures are not germane to neuronal networks. Indeed, cortical networks tend to rather display complex network topologies~\cite{bosking-zhang-etal:97}. Typical cortical networks tend favor local connectivity: they present a short path length (associated with global efficacy of information transfer), high clustering (associated to resilience to noise) which are rather compatible with small-world topologies and that ensure important function (see~\cite{bassett2006small}, and ~\cite{bullmore2009complex} for a review).  Moreover, some experimental studies tend to relate typical connectivity patterns with collective qualitative properties of the networks in physiological situations~\cite{gray1989oscillatory}, and in particular in relationship with the emergence of synchronized activity. The question we may ask here is whether such random architectures, in which neurons connect to their nearest neighbors with higher probability than to more remote cells, have qualitative properties different from fully connected networks. 

The topic of this paper is precisely to investigate the role of network topology in the macroscopic or mesoscopic activity of cortical networks. From a mathematical viewpoint, heterogeneous connections break down the interchangeability assumption usually instrumental in order to prove mean-field limits (see e.g.~\cite{sznitman:89}). However, the classical coupling method readily extends, as we show here, to networks with specific random topologies. In detail, a weak notion of exchangeability under a certain probability law (that of the connectivity weights and delays) is enough. We will address here both annealed (i.e. averaged over all possible connectivity patterns) and quenched convergence along subsequences, of networks with random architectures and random delays in a general setting encompassing the classical models of Hodgkin-Huxley and Fitzhugh-Nagumo neurons~\cite{ermentrout-terman:10}. In order to uncover the role of random connectivities in the qualitative behavior of the network, we will instantiate a simple model, the WC firing-rate model with noise as a benchmark of single neurons behaviors. This model has the interest (see~\cite{touboul-hermann-faugeras:11}) to have Gaussian solutions whose mean and standard deviation satisfy a dynamical system which will be analyzed using the bifurcation theory. The rigorous analysis of this model will lead us to conclude that in our models, random connectivities affect the network only when these are correlated with the delays (which is the case in neuronal network models since the connectivity probability, as well as the interaction delays are functions of the distance between two cells), and that these topologies govern the response of the network. 

The paper is organized as follows. We start by introducing in section~\ref{sec:setting} the formalism and the network under consideration. In section~\ref{sec:MainResults} we present the main theoretical results for finite-populations networks on which our developments are based. Appendix~\ref{append:InfiniteDimensions} extends these results to neural fields where the number of different neuronal populations tends to infinity. The proof of these results are extension of previous works~\cite{touboulNeuralFieldsDynamics:11,toubouldelays:12}, and are therefore postponed in section~\ref{sec:Proofs}.  Section~\ref{sec:Applications} is devoted to the qualitative analysis of the nature of the solutions in the case of the firing-rate model, and exhibits the relationship between network topology and macroscopic dynamics. 

\section{Setting of the problem}\label{sec:setting}

We now introduce the mathematical formalism used throughout the paper. We work in a complete probability space $(\Omega,\F,\PP)$ satisfying the usual conditions. We will analyze the dynamics of a neuronal network model composed of $N$ neurons, in  an abstract setting valid for most usual models used in computational neuroscience such as the Hodgkin-Huxley \cite{hodgkin-huxley:52} or Fitzhugh-Nagumo \cite{fitzhugh:55} models. In this model, the state of each neuron $i$ is described by a $d-$dimensional variable $X^{i,N}\in E$ (typically in $E\subset\RR^d$) corresponding to the membrane potential, ionic concentration and gated channels (see e.g. \cite{ermentrout-terman:10b}).

The networks are composed of $N$ neurons falling into $P(N)$ populations labeled $\alpha\in\{1,\ldots,P(N)\}$ and composed of $N_{\alpha}$ neurons, and the convention $\alpha=p(i)$ defines the population neuron $i$ belongs to. The level of description chosen governs the choice of the asymptotic regime analyzed. Here, we will consider two main different cases: 
\begin{itemize}
	\item The \emph{macroscopic scale} where neurons gather into a few populations  $P(N)=P$ fixed, corresponding to coarse-grained descriptions of neural activity, generally called in the neuroscience domain neural mass models~\cite{jansen-rit:95}
	\item The \emph{mesoscopic scale}, or neural-field limit, where the number of populations tends to infinity and the area described covers a continuous piece of cortex $\Gamma\subset\RR^p$ with $p\in \N^*$. This description correspond to finer scale descriptions at which averaging effects occur, but fine enough to resolve the spatial structure of the cortex~\cite{touboulNeuralfields:11}.
\end{itemize} 

In each population, neurons have different intrinsic properties, receive different inputs and present a specific connectivity map with neurons in the other populations. Macroscopic or mesoscopic behavior correspond to the network's properties when the number of neurons in each population tends to infinity. This limit will be denoted with a slight abuse of notation $N\to \infty$. To fix ideas, we make the following assumption in the macroscopic scale case:
\begin{itshape}
	\renewcommand{\theenumi}{(H\arabic{enumi})}

\begin{enumerate}
		\setcounter{enumi}{-1}
\item\label{assump:H0} There exists a sequence of positive real numbers $r_1,\ldots,r_{P} \in (0,1)^P$ with $\sum_{\alpha=1}^P r_{\alpha}=1$ such that for all $\alpha\in\{1,\ldots,P\}$, $$N_\alpha/N\longrightarrow r_\alpha,$$ when $N\rightarrow\infty$.
\end{enumerate}
\end{itshape}
In other words, we are assuming that the fraction of neurons belonging to a given population remains non trivial in the limit $N\to\infty$.

The evolution state $X^{i,N}_t$ of neuron $i$ in the population $\alpha\in\{1,\cdots,P\}$ is governed by a stochastic differential equation. The intrinsic dynamics of the neuron is governed by a function $f_{\alpha}:\RR_+\times E\mapsto E$. This evolution is stochastic, driven by independent $m$-dimensional Brownian motions $(W^i_t)$ through a diffusion coefficient $g_\alpha:\RR_+\times E\mapsto \R^{d\times m}$. The neuron $i$ receives inputs from other neurons in the network, which affect its state through an interaction function $b_{\alpha\gamma}:\R\times E\times E\mapsto E$ depending on
	\begin{itemize}
		\item The synaptic weight $w_{ij}\in\RR$ between neurons $i$ in population $\alpha$ and $j$ in population $\gamma$ controlling the topology of the network: these are zero when there is no connection between $i$ and $j$, positive when the connection is excitatory and negative when inhibitory. 
		\item the state of both neurons $i$ and $j$.
	\end{itemize}
These interactions take place after a delay $\tau_{ij}>0$.

The dynamics of neuron $i$ in population $\alpha$ is given by:
\begin{multline}
 dX_t^{i,N}=\Big(f_\alpha(t,X_t^{i,N})+\sum_{\gamma=1}^{P(N)}\sum_{p(j)=\gamma}\frac1{N_\gamma}b_{\alpha\gamma}(w_{ij},X_t^{i,N},X_{t-\tau_{ij}}^{j,N})\Big)dt\\
+g_\alpha(t,X_t^{i,N})\cdot dW_t^i,
 \label{eq:Network}
\end{multline}
under the assumption that $b(0,x,y)=0$ and the fact that the synaptic weight $w_{ij}$ is assumed zero when no link from $j$ to $i$. In these notations, the architecture of the network is completely integrated in the choice of the synaptic coefficients $w_{ij}$. In our purpose to analyze networks on random graphs, we will consider that the synaptic weights $w_{ij}$ and delays $\tau_{ij}$ are non-negative random variables, drawn in a distinct probability space $(\Omega',\F',\mathcal{P})$ at the beginning of the evolution and frozen. We generically denote by $\mathbb{E}$ the expectation with respect to the processes (i.e., under the probability $\mathbb{P}$)  and by $\E$ the expectation of random variables or processes with respect to the environment (i.e. under $\mathcal{P}$). One realization of these weights corresponds to one network with prescribed architecture. In realistic settings, both connectivity weight and delay are related to the distance between the cells, and therefore are generally correlated. A specific choice relevant to biology is discussed in section~\ref{sec:Applications}, in which connectivity probability as well as delays are deterministic functions of the random respective locations of the cells (yielding a specific strong correlation between the two variables). 

While the random variables $w_{ij}$ and $\tau_{ij}$ are correlated, an important hypothesis is that for fixed $i$,  sequences $(\tau_{ij}, j=1\cdots N)$ and $(w_{ij}, j=1\cdots N)$ are considered independent and identically distributed population-wise, i.e. they have the same distribution for all $j$ belonging to a given population\footnote{Note that the whole sequence of weights $(w_{ij}; 1\leq i,j\leq N)$ as well as the delays $(\tau_{ij}; 1\leq i,j\leq N)$ might be correlated. When these are related to the distance $r_{ij}$ between $i$ and $j$, correlations may arise from symmetry ($r_{ij}=r_{ji}$) or triangular inequality $r_{ij}\leq r_{ik}+r_{kj}$. The independence assumption is nevertheless valid in that setting provided that the locations of the different cells are independent and identically distributed random variables}. For fixed $i\in\{1,\cdots,N\}$, we can therefore denote $\Lambda_{ip(j)}$ the distribution of the variables $(w_{ij},\tau_{ij})$. 

The piece of cortex considered will be said invariant by translations if the joint distribution of weights and delays $(w_{ij},\tau_{ij})$ for $p(j)=\gamma$ does not depend on the specific neuron $i$ considered but only on the population $\alpha$ the neuron $i$ belongs to\footnote{The term \emph{invariant by translation} is chosen in reference to random variables $\tau_{ij}$ and $w_{ij}$ function of the distance $r_{ij}$ between neuron $i$ and $j$: this distance is independent of the particular choice of neuron $i$ (and of its location) if the neural field is invariant by translation in the usual sense}. In that case, we will denote $\Lambda_{\alpha\gamma}$ the joint law of weights and delays. In the \emph{general case}, we assume that the laws $\Lambda_{i\gamma}$ are independently drawn from a distribution of measures centered at a specific one $\Lambda_{\alpha\gamma}$. For instance, when delays and connectivity depend on the distance between cells, the distribution $\Lambda_{i\gamma}$ depends on the position $r_i$ of neuron $i$. If cells of population $\alpha$ are distributed on a space $D$ with density $p$, and the weights and delays have a density $\lambda_{r_i}(s,t)$, $\Lambda_{\alpha\gamma}$ is the law with density $\int_{D} \lambda_r(s,t)dp(r)$. 

Let us denote by $\tau$ the maximal possible delay $\tau_{ij}$ which we assume finite\footnote{This is always the case when considering bounded neural fields.}. Equations~\eqref{eq:Network} are stochastic differential equations on the infinite-dimensional space of functions $C([-\tau,0],E)$ (i.e. on the variable $\tilde X_t = (X_s, s\in[t-\tau,t])$, see e.g. \cite{da-prato:92,mao:08}).

Finally, we consider that the network has chaotic initial states, in the sense that they have independent and population-wise identically distributed initial conditions. In detail, we denote $C_\tau=C([-\tau,0],E^P)$ and set $(\zeta_0^\alpha(t))\in C_\tau$ a
stochastic process with independent components. Chaotic initial condition on the network consists in setting independent initial condition for all neurons, with distribution for neurons of population $\alpha$ equal to that of $\zeta_0^\alpha$.

In what follows, we note $\mathcal M^2\big(C([-\tau, 0],E^N)\big)$ the space of square integrable stochastic processes on $[-\tau, 0]$ with values in $E^N$, $\mathcal M(\mathcal C)$ the set of probability distributions on $\mathcal C$ the set continuous functions $[-\tau,T]\mapsto E^P$, and $\mathcal M^2(\mathcal C)$ the space of square-integrable
processes. 

\section{Main results}\label{sec:MainResults}

In this section, we state and discuss the main mathematical results on the convergence of the above described process as the network size goes to infinity. Interestingly, even if the network considered has a complex random topology in which connectivity map as well as delays are correlated, methods developed in the case of fully connected architectures~\cite{toubouldelays:12,touboulNeuralfields:11} extend to this more complex case. Proofs are provided for completeness in section~\ref{sec:Proofs}.

Let us first state the following proposition ensuring well-posedness of the network system:

\begin{prop}\label{prop1}
 Let $X_0\in\mathcal M^2(C([-\tau,0],E^N))$ an initial condition of the network system. For any $(\alpha,\gamma)\in\{1,\ldots,P(N)\}^2$, assume that:
	\renewcommand{\theenumi}{(H\arabic{enumi})}
 \begin{enumerate}
 \item $f_\alpha$ and $g_\alpha$ are uniformly in time Lipschitz-continuous functions with respect to their second variable.
 \item For almost all $w\in\RR$, $b_{\alpha\gamma}(w,\cdot,\cdot)$ is $L_{\alpha\gamma}$-Lipschitz-continuous with respect of both variables.
 \item There exists functions $\bar K_{\alpha\gamma} : \R\mapsto \RR^+$ such that for any $(\alpha,\gamma)\in\{1,\cdots,P(N)\}^2$, $$|b_{\alpha\gamma}(w,x,y)|^2\leq \bar K_{\alpha\gamma}(w)\qquad\mbox{and}\qquad \E[\bar{K}_{\alpha\gamma}(w)]\leq\bar k<\infty.$$
 \item The drift and diffusion functions satisfy the monotone growth condition: there exists a positive constant $K$ depending on $f$ and $g$ such that: $$x^Tf_{\alpha}(t,x)+\frac12|g_\alpha(t,x)|^2\leq K(1+|x|^2).$$
\end{enumerate}
Then for almost all realization of the synaptic weights $w_{ij}\in\R$ and the delays $\tau_{ij}\in[-\tau,0]$, we have existence and uniqueness of solutions to the network equations \eqref{eq:Network}.
\end{prop}

This property results from the application of standard theory of stochastic delayed differential equations. We provide a sense of the proof in  section~\ref{sec:Proofs}: the details of the proof of this elementary proposition will largely simplify the analysis of the limit equations.\\

When the number of neurons goes to infinity (under assumption~\ref{assump:H0}) then
\begin{itemize}
 \item for almost any realization of the transmission delays $\tau_{ij}$ and synaptic weights $w_{ij}$ in the translation-invariant case or
 \item averaged across all realizations of the disorder in the general case,
\end{itemize}
the propagation of chaos property holds: {\it if the initial conditions are chaotic, then the states of a finite number of neurons are independent for all times when $N\rightarrow\infty$.} Their law is given by a nonlinear McKean-Vlasov equation that depends on the neural population they belong to. Similar results hold for mesoscopic limits of neural field models, i.e. in situations in which the number of populations $P(N)$ diverges as $N\to\infty$. In this case, the notion of solution is much more complex, as one obtains a process depending on space but which is not measurable with respect to the spatial variable. These questions, addressed in~\cite{touboulNeuralfields:11}, will be briefly discussed in our context in appendix~\ref{append:InfiniteDimensions}.

In both cases, the proof of the convergence and propagation of chaos will use the powerful coupling method (see \cite{sznitman:89}). The proof is in two steps: (i) we prove that the limit equation (see equation~\eqref{eq:MeanField1} below) has an unique solution, and (ii) that the law of $X^{i,N}_t$ converges towards the law of \eqref{eq:MeanField1}\footnote{More precisely, taking a finite set of neurons $\{i_1,\ldots,i_k\}$ the law of the process $(X_t^{i_1,N},\ldots,X_t^{i_1,N},t\in[-\tau,T])$ converge in probability towards a vector $(\bar X_t^{i_1},\ldots,\bar X_t^{i_1},t\in[-\tau,T])$, where the processes $\bar X^l$ are independent and have the law of $X^{p(i_l)}$ given by \eqref{eq:MeanField1}.}.

\subsection{Randomly connected neural mass models}

Let $P(N)=P$ be fixed and independent of $N$. In this case, we will show that the network equation converges (in a sense to be defined in each sub case) towards the solution of a well-posed McKean-Vlasov equation given by:
\begin{multline}
 d\bar{X}^\alpha_t=f_\alpha(t,\bar{X}^\alpha_t)\,dt+g_\alpha(t,\bar{X}^\alpha_t)\cdot dW^\alpha_t\\
+\Big(\sum_{\gamma=1}^P\int_{-\tau}^0\int_{\R}\EE_{\bar Y}\big[b_{\alpha\gamma}\left(w,\bar{X}^\alpha_t,\bar Y_{t+s}^\gamma\right)\big]d\Lambda_{\alpha\gamma}(s,w)\Big)dt,
\label{eq:MeanField1}
\end{multline}
where $\bar Y$ is a process independent of $\bar X$ that has the same law, $\EE_{\bar Y}$ the expectation under the law of $\bar Y$, and $W_t^\alpha$ are independent adapted standard Brownian motions of dimension $d\times m$. Denoting by $m^\gamma_t(dx)$ the law of $\bar X^\gamma_t$ the equation \eqref{eq:MeanField1} is nothing but
\begin{multline}
 d\bar{X}^\alpha_t=f_\alpha(t,\bar{X}^\alpha_t)\,dt+g_\alpha(t,\bar{X}^\alpha_t)\cdot dW^\alpha_t\\+\Big(\sum_{\gamma=1}^P\int_{-\tau}^0\int_{\R}\int_E\big[b_{\alpha\gamma}\left(w,\bar{X}^\alpha_t,y\right)\big]m_{t+s}^\gamma(dy)d\Lambda_{\alpha\gamma}(s,w)\Big)dt,
\label{eq:MeanField2}
\end{multline}



The hypotheses made in Proposition~\ref{prop1} also ensure existence and uniqueness of solutions as we now state in the following:
\begin{theo}
Under the hypotheses of Proposition~\ref{prop1} and for any $\zeta_0\in\mathcal M(C([-\tau,0],E^P))$ a square integrable process, the mean-field equations \eqref{eq:MeanField2} with initial condition $\zeta_0$ have a unique strong solution on $[-\tau,T]$ for any time horizon $T>0$.
 \label{theo1}
\end{theo}

In order to demonstrate the convergence of the network equation and the propagation of chaos when the number of neurons goes to infinity, we use Dobrushin's coupling approach \cite{dobrushin:70,sznitman:84a,sznitman:89,tanaka:78} in the same fashion as done in \cite{toubouldelays:12,touboulNeuralfields:11} in the context of neurosciences, the only difference being the random environment nature of the network equation related to the random structure of the synaptic coefficients. 

\subsection{Quenched convergence and propagation of chaos in the translation invariant case}

The translation invariant case correspond to the situation where the laws $\Lambda_{i\gamma}$ for $i$ such that $p(i)=\alpha$ are identical and only depend on $\alpha$.

Let $i\in\mathbb N$ such that $p(i)=\alpha$. We define the process $\bar X^i$ solution of \eqref{eq:MeanField1}, driven by the Brownian motions $(W^i_t)$ that governs $X^i$, and having the same initial condition as neuron $i$ in the network, $\zeta_0^i\in\mathcal M^2(\mathcal C)$:
\begin{equation}
\begin{cases}
  d\bar{X}^i_t&= f_\alpha(t,\bar X^i_t)dt+g_\alpha(t,\bar X^i_t)\cdot dW_t^i\\
&\displaystyle\qquad \qquad+\Big(\sum_{\gamma=1}^P\int_{-\tau}^0\int_{\R}\EE_Z\big[b_{\alpha\gamma}\big(w,\bar X^i_t,Z_{t+s}^\gamma\big)\big]d\Lambda_{\alpha\gamma}(s,w)\Big)dt , \quad t\geq0\\
  \bar X^i_t&=\zeta_0^i(t),\, t\in[-\tau,0].
\end{cases}
 \label{eq:Quenched}
\end{equation}
By definition, the processes $(Z_t^1,\ldots,Z_t^P)$ are a collection of processes independent of $(\bar X_t^i)_{i=1,\ldots,N}$ and have the distribution $m^1_t\otimes\cdots\otimes m^P_t$, where $m^\alpha_t$ is the probability distribution of $\bar X^\alpha_t$ (unique solution of the equation~\eqref{eq:MeanField1}).

Theorem~\ref{theo1} ensures well posedness of these equations, and therefore $(\bar X^i_t)_{i\in\N}$ constitute a sequence of independent processes with law $\bar X^{p(i)}$. 

\begin{theo}[Quenched Convergence]\label{theo2}
Under assumptions (H1)-(H4) and chaotic initial conditions in $\mathcal M^2(\mathcal C)$. The process $(X^{i,N}_t,-\tau\leq t\leq T)$ for $i\in\N$ fixed, solution of the network equations \eqref{eq:Network}, converges almost surely towards the process $(\bar X^i_t,-\tau\leq t\leq T)$ solution of the mean-field equations \eqref{eq:Quenched}. This implies in particular convergence in law of the process $(X^{i,N}_t,-\tau\leq t\leq T)$ towards $(\bar X^\alpha_t,-\tau\leq t\leq T)$ solution of the mean-field equations \eqref{eq:MeanField1}.
\end{theo}

%

\subsection{Annealed convergence and propagation of chaos in the general case}

We now turn our attention to the case of non-translation invariant networks where the law of delays and synaptic weights depend on the index of neuron $i$ in population $\alpha$. In this case we will see that the propagation of chaos property remains valid as well as convergence to the mean-field equations~\eqref{eq:MeanField1}, no more for almost all realization of the disorder, but in average across all possible configurations. Denoting $\mathcal E_i$ the expectation over all possible distributions $\Lambda_{i\gamma}$, we have:

\begin{theo}[Annealed convergence in the general case]\label{theo3}
We assume that (H1)-(H4) are valid and that network initial conditions are chaotic in $\mathcal M^2(\mathcal C)$, and that the interaction does not depend on the postsynaptic neuron state (i.e., $b(w,x,y)=\ell(w,y)$). Let us fix $i\in\N$, then the law of process $(X^{i,N}_t,\;-\tau\leq t\leq T)$ solution to the network equations \eqref{eq:Network} averaged over all the possibles realizations of the disorder, converge almost surely towards the process $(\bar X^i_t,\;-\tau\leq t\leq T)$ solution to the mean field equations \eqref{eq:MeanField1}. This implies in particular the convergence in law of $(\mathcal E_i[X^{i,N}_t],\;-\tau\leq t\leq T)$ towards $(\bar X^\alpha_t,\;-\tau\leq t\leq T)$ solution of the mean field equations \eqref{eq:MeanField1}.
\end{theo}

Extensions to the spatially extended neural field case are discussed in Appendix~\ref{Append:NeuralField}.

\section{Application: dynamics of the firing-rate model with random connectivity}\label{sec:Applications}

In the previous section, we derived limit equations for networks with random connectivities and synaptic weights. The motivation of these mathematical developments is to understand the role of specific connectivity and delays patterns arising in plausible neuronal networks. More precisely, it is known that anatomical properties of neuronal networks affect both connectivities and delays, and we will specifically consider the two following facts:
\begin{itemize}
	\item Neurons connect preferentially to those anatomically close.
	\item Delays are proportional to the distance between cells.
\end{itemize}

At the level of generality of the previous sections, we obtained very complex equations, from which it is very hard to uncover the role of random architectures. However, as we already showed in previous works~\cite{touboul-hermann-faugeras:11}, a particularly suitable framework to solve these questions is provided by the classical firing-rate model. In that case, we showed in different contexts that the solution to the mean-field equations is Gaussian, whose mean and standard deviation are solution of simpler dynamical system. 



\subsection{Reduction to distributed delays differential equations} 

In the firing-rate model, the intrinsic dynamics of each neuron is given by $$f_\alpha(t,x)=-x/\theta_\alpha+I_\alpha(t),$$ where $I_\alpha(t)$ is the external input of the system, and the diffusion function $g_\alpha(t,x)=\lambda_\alpha$ is constant. The interaction only depends in a nonlinear transform of the membrane potential of the pre-synaptic neuron multiplied by the synaptic weight: $b_{\alpha\gamma}(w,x,y)=J_{\alpha\gamma}(w)S(y)$. We also assume, in order to satisfy the assumptions of the Theorems~\ref{theo2} and \ref{theo3}, that the functions $J_{\alpha\gamma}\in L^{\infty}(\mathcal \RR)$ and $S\in W^{1,\infty}(E^d)$. Therefore, when considering the delays and the synaptic weights only depending on $p(i)$,  we have propagation of chaos and almost sure convergence (quenched) towards the mean-field equations:
\begin{eqnarray}
 d\bar{X}^\alpha_t&=&\Big(-\frac{\bar{X}^\alpha_t}{\theta_\alpha}+I_\alpha(t)+\sum_{\gamma=1}^P\int_{-\tau}^0\int_\R J_{\alpha\gamma}(w)\,\EE_{{Y}}\big[S({Y}_{t+s}^\gamma)\big]d\Lambda_{\alpha\gamma}(s,w)\Big)\,dt\nonumber \\
&&\qquad +\lambda_\alpha dW^\alpha_t,
\label{eq:FiringRate}
\end{eqnarray}
and in the general case, the same result holds in an averaged sense. 
\begin{remark}
\emph{	Let us note that if the synaptic weights and the delays are independent, it is very easy to see that the network converges towards an effective mean-field equation where the disorder in the connectivity weights disappears and the mean-field equation obtained reduces to
	\begin{equation*}
	 d\bar{X}^\alpha_t =\Big(-\frac{\bar{X}^\alpha_t}{\theta_\alpha}+I_\alpha(t)+\sum_{\gamma=1}^P\bar{J}_{\alpha\gamma}\int_{-\tau}^0 \EE_{{Y}}\big[S({Y}_{t+s}^\gamma)\big]d\rho_{\alpha\gamma}(s)\Big)\,dt +\lambda_\alpha dW^\alpha_t,
	\end{equation*}
	where $\rho_{\alpha\gamma}$ is the marginal density of delays of $\Lambda_{\alpha\gamma}$ and $\bar{J}_{\alpha\gamma}$ is the averaged synaptic weight. This is exactly the same equation as would arise from a non-disordered network equation where all connectivity weights are deterministic: $J_{ij}=\bar{J}_{\alpha\gamma}/N_{\gamma}$. Therefore, the architecture plays a role in the dynamics only when the synaptic weights and the delays are correlated, as is the case of the cortex. }
\end{remark}

We will therefore focus on more realistic models where delays and connectivity weights are correlated. It is very easy to see, integrating equation~\eqref{eq:FiringRate}, that the solution satisfies the implicit equation:
\begin{eqnarray*}
&& \bar X^\alpha_t=\bar X^\alpha_0e^{-t/\theta_\alpha}+\int_0^te^{-(t-s)/\theta_\alpha}\Big(-\frac{\bar{X}^\alpha_s}{\theta_\alpha}+I_\alpha(s)\\
&&\qquad+\sum_{\gamma=1}^P\int_{-\tau}^0\int_\R J_{\alpha\gamma}(w)\,\EE_{\bar{Y}}\big[S({Y}_{s+r}^\gamma)\big]d\Lambda_{\alpha\gamma}(r,w)\Big)\,ds+\int_0^te^{-(t-s)/\theta_\alpha}\lambda_\alpha dW^\alpha_s
\end{eqnarray*}
which is composed of Gaussian terms and the initial condition $\bar X^\alpha_0e^{-t/\theta_\alpha}$ vanishing at an exponential rate.
Therefore, when the initial conditions are Gaussian processes\footnote{If the initial condition is not Gaussian, the solution to the mean-field equation will nevertheless be attracted exponentially fast towards the Gaussian solution described.}, the solution is also Gaussian with mean $u_\alpha$ and variance $v_\alpha$. Taking expectation and covariance we get that the mean and the variance of the solution satisfy the following well-posed system of delayed differential equations:
\begin{equation}
 \left\{\begin{array}{l}
  \displaystyle\dot u_\alpha=- u_\alpha/\theta_\alpha+\sum_{\gamma=1}^P \int_{-\tau}^0\int_\R J_{\alpha\gamma}(w)\,\EE_{{Y}}\left[S({Y}_{t+s}^\gamma)\right]d\Lambda_{\alpha\gamma}(s,w)\\
  \displaystyle\dot v_\alpha = -2v_\alpha/\theta_\alpha+\lambda_\alpha^2.
 \end{array}\right.
\end{equation}

In the firing-rate case, we hence have an important reduction of complexity. This simpler form allows us to use bifurcation theory in order to understand the role of the parameters on the qualitative properties of the solutions. This theory has been widely used in neuroscience in order to uncover, in single cells models, the emergence of periodic spiking or bursting~\cite{ermentrout-terman:10b}, and for heuristic macroscopic models, formation of patterns of activity~\cite{bressloff:12} or visual hallucinations~\cite{ermentrout-cowan:79}. Here, the theory of delayed differential equations (see e.g.~\cite{ermentrout-cowan:79}) allows us to uncover the role of the randomness of the architecture and delays in shaping the collective  behavior of the network. In order to analyze this dependence, we consider the system in the absence of external input $I=0$ and $$S(x):=\frac1{\sqrt{2\pi}}\int_0^xe^{-s^2/2}\,ds,$$ which has the property that a simple change of variables yields (see~\cite[Appendix A]{touboul-hermann-faugeras:11}):
\begin{equation}\nonumber
\EE_{ Y}[S( Y_{t}^{\gamma})]=\EE_{ Y}[S( Y_{t}^{\gamma})]=S\Big(\frac{u_\gamma(t)}{\sqrt{1+v_\gamma(t)}}\Big)
\end{equation}
In that simplified case, a stationary solution of the system is given by $(u_{\alpha}^\ast,v_{\alpha}^\ast)=(0,\lambda_\alpha^2\theta_\alpha/2)$. The solution to the variance equation is $$v_\alpha(t)=\frac 1 2(\lambda_\alpha^2\theta_\alpha+e^{-2t/\theta_\alpha})=v_\alpha^\ast+\frac 1 2 e^{-2t/\theta_\alpha},$$ then the stability of the fixed point only depends on the delayed linear equation to the mean, which is:
\begin{equation}\nonumber
\dot u_\alpha(t)=-\frac{u_\alpha(t)}{\theta_\alpha}+\sum_{\gamma=1}^P\int_{-\tau}^0\int_\R J_{\alpha\gamma}(w)\frac 1{\sqrt{2\pi(1+v_\gamma^\ast)}}u_\gamma(t+s)\,d\Lambda_{\alpha\gamma}(s,w).
\end{equation}
If only one population is considered, then dropping the index for the population lead us to:
\begin{equation}\label{eq:NotTheta}
\dot u(t)=-\frac{u(t)}{\theta}+\int_{-\tau}^0\int_\R J(w)\frac 1{\sqrt{2\pi(1+v^*)}}u(t+s)\,d\Lambda(s,w).
\end{equation}
The stability of the fixed point only depends on the dispersion relationship:
\begin{equation}\label{eq:DispRel}
 \xi=-\frac1\theta+\frac{1}{\sqrt{2\pi(1+v^*)}}\int_{-\tau}^0\int_\R J(w)\,e^{\xi s}\,d\Lambda_{\alpha\gamma}(s,w),
\end{equation}
which is nothing more that looking for solutions of the form $u=\exp(\xi t)$ in~\eqref{eq:NotTheta}.

The solutions of this equations are the characteristic exponents of the system, and relate directly the stability of the fixed point considered. If all characteristic exponents have negative real part, the equilibrium is asymptotically exponentially stable, but if there exists a characteristic exponent with strictly positive real part, the equilibrium is unstable. Turing-Hopf bifurcations occur when the system has a pair of complex conjugate characteristic exponents with non-zero imaginary part crossing the imaginary axis.

\subsection{Small-world type model and correlated delays} As we stated before one interesting situation arising in neuroscience is the case where synaptic weights and the delays are function of the distance between neurons. Without loss of generality, we assume the signal transmission speed is unitary, then the delay $\tau_{ij}$ between the neuron $i$ at location $r_i$ and a neuron $j$ at location $r_j$ is simply modeled by $$\tau_{ij}=|r_i-r_j|+\tau_s,$$ where $\tau_s$ is the minimum value corresponding to the transmission of the information at the synapse. We further assume that the synaptic links are drawn according to a Bernoulli random variable:
\begin{equation}\nonumber
w_{ij}=\begin{cases}
 1&\text{with probability } b(|r_i-r_j|):=e^{-\beta|r_i-r_j|}\\
 0&\text{with probability }1-b(|r_i-r_j|),
\end{cases}
\end{equation}
with $\beta>0$. The synaptic weights are given by $J(w_{ij})$ with 
\[J(x)=\begin{cases}
	\bar{J} & \text{if }x=1\\
	0 & \text{if }x=0
\end{cases}.\]

%


In this model, the total connectivity level of the system decreases when $\beta$ is increased. When neurons are uniformly distributed in the interval $[0,a]$, the averaged law density can be easily computed and is given by: $$dp(r)=\left(\frac{2}{a}-\frac{2r}{a^2}\right)\,dr,$$ and thanks to conditional expectation we find that \eqref{eq:DispRel} is nothing but
\begin{eqnarray*}
  \xi&=&-\frac1\theta+\frac{1}{\sqrt{2\pi(1+v^*)}}\E\big[\E\big[J(w)e^{\xi u}\big|r\big]]\\
  &=&-\frac1\theta+\frac{1}{\sqrt{2\pi(1+v^*)}}\E\big[\E\big[J(w)\big|r\big]e^{-\xi(\tau_s+r)}]\\
  &=&-\frac1\theta+\frac{\bar Je^{-\xi\tau_s}} {\sqrt{2\pi(1+v^*)}}\int_0^ae^{-(\beta+\xi)r}\left(\frac{2}{a}-\frac{2r}{a^2}\right)\,dr.
\end{eqnarray*}
Turing bifurcations arise for parameters such that there exists a purely imaginary characteristic root (solution of the above equation) $\xi=i\omega$. These occur when one can find $\omega>0$ such that:
\begin{multline}\label{eq:Hopf1}
 i\omega=-\frac1\theta+\frac{2\bar J}{\sqrt{2\pi(1+v^*)}}\times\\\frac{1}{a(\beta+i\omega)}\left(1-\frac1{a(\beta+i\omega)}+\frac{e^{-a(\beta+i\omega)}}{a(\beta+i\omega)}\right)e^{-i\omega\tau_s}.
\end{multline}
Since \eqref{eq:Hopf1} depend on many parameters, in order to understand the solutions we study the system decoupling the size of the neural field with respect to the connectivity parameter $\beta$ and the size $a$.

\subsubsection{The effect of the extension of the neural field.}

We first fix $\beta>0$ and make the change of variables $\Omega=a\omega$, $B=a\beta$. Defining $$Z(\Omega,B)=\frac{2\bar J}{\sqrt{2\pi(1+v^*)}}\frac1{B+i\Omega}\left(1-\frac1{B+i\Omega}+\frac{e^{-(B+i\Omega)}}{B+i\Omega}\right),$$ then \eqref{eq:Hopf1} is reduced to solve the system
\begin{equation}\label{eq:Paramao:02}
\begin{cases}
 a^2=\Omega^2\left(|Z(B,\Omega)|^2-\frac{1}{\theta^2}\right)^{-1},\\
 \tau_s=\left(Arg(Z(\Omega,B))-Arg\left(1+\frac{i\Omega}{a}\right)+2k\pi\right)\frac{a}\Omega,\\
 B=\beta a
\end{cases}
\end{equation}
which can be seen as a intersection of two surfaces in the space $(a,B,\tau_s)$:
\begin{eqnarray*}\nonumber
 S_1:\left\{\begin{array}{ccc}
  \RR\times\RR_+&\rightarrow&\RR^3\\
 (\Omega,B)&\mapsto& (a(\Omega,B),B,\tau_s(\Omega,B))
 \end{array}\right. &\quad&
 S_2:\left\{\begin{array}{ccc}
  \RR_+\times\RR&\rightarrow&\RR^3\\
 (a,\tau_s)&\mapsto& (a,\beta a,\tau_s)
 \end{array} \right.,
\end{eqnarray*}
where $a(\Omega,B)$ and $\tau_s(\Omega,B)$ are the solutions of \eqref{eq:Paramao:02} for $B$ given. We obtain a sequence of Turing-Hopf bifurcations indexed by $k$, and the first bifurcation is responsible for oscillations appearing in the system.

\begin{figure}
\centering
 \subfigure[$\beta=0.1$ fixed]{
  \includegraphics[width=0.45\textwidth]{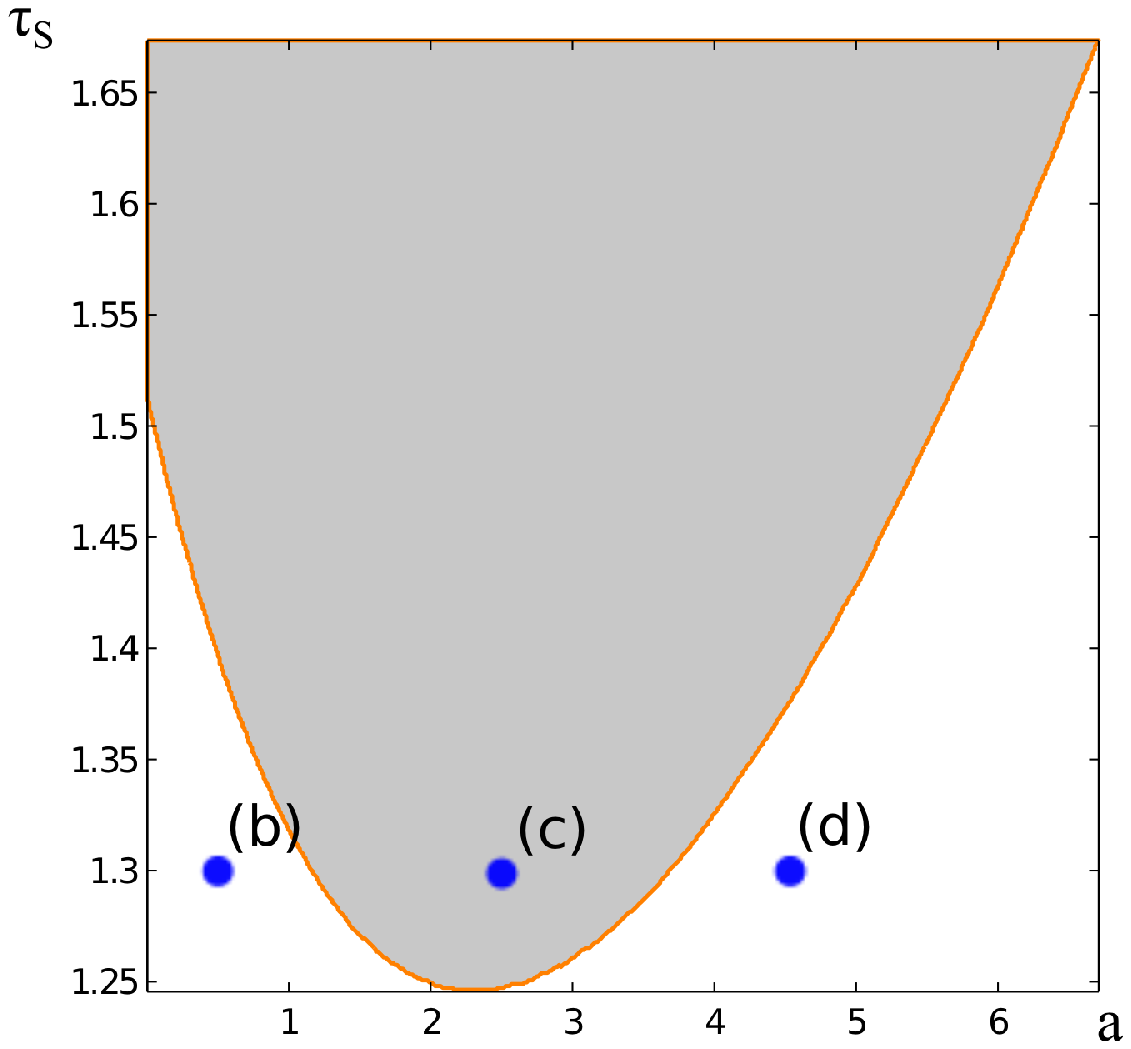}
   \label{fig:hopf3d}
 }
 \subfigure[$a=0.5$]{
  \includegraphics[width=0.45\textwidth]{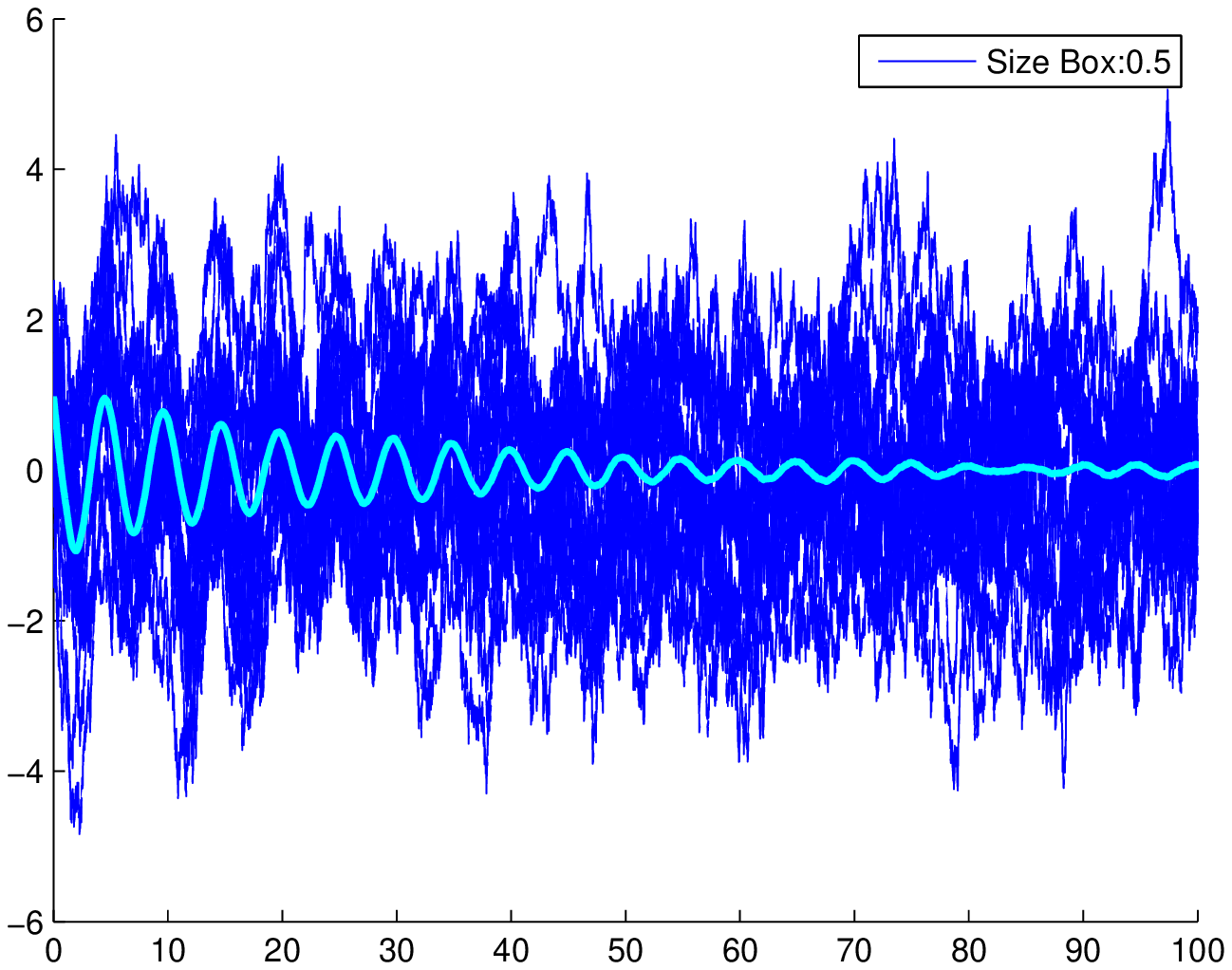}
   \label{fig:osc1a0_1}
 }
  \centering
 \subfigure[$a=2.5$]{
  \includegraphics[width=0.45\textwidth]{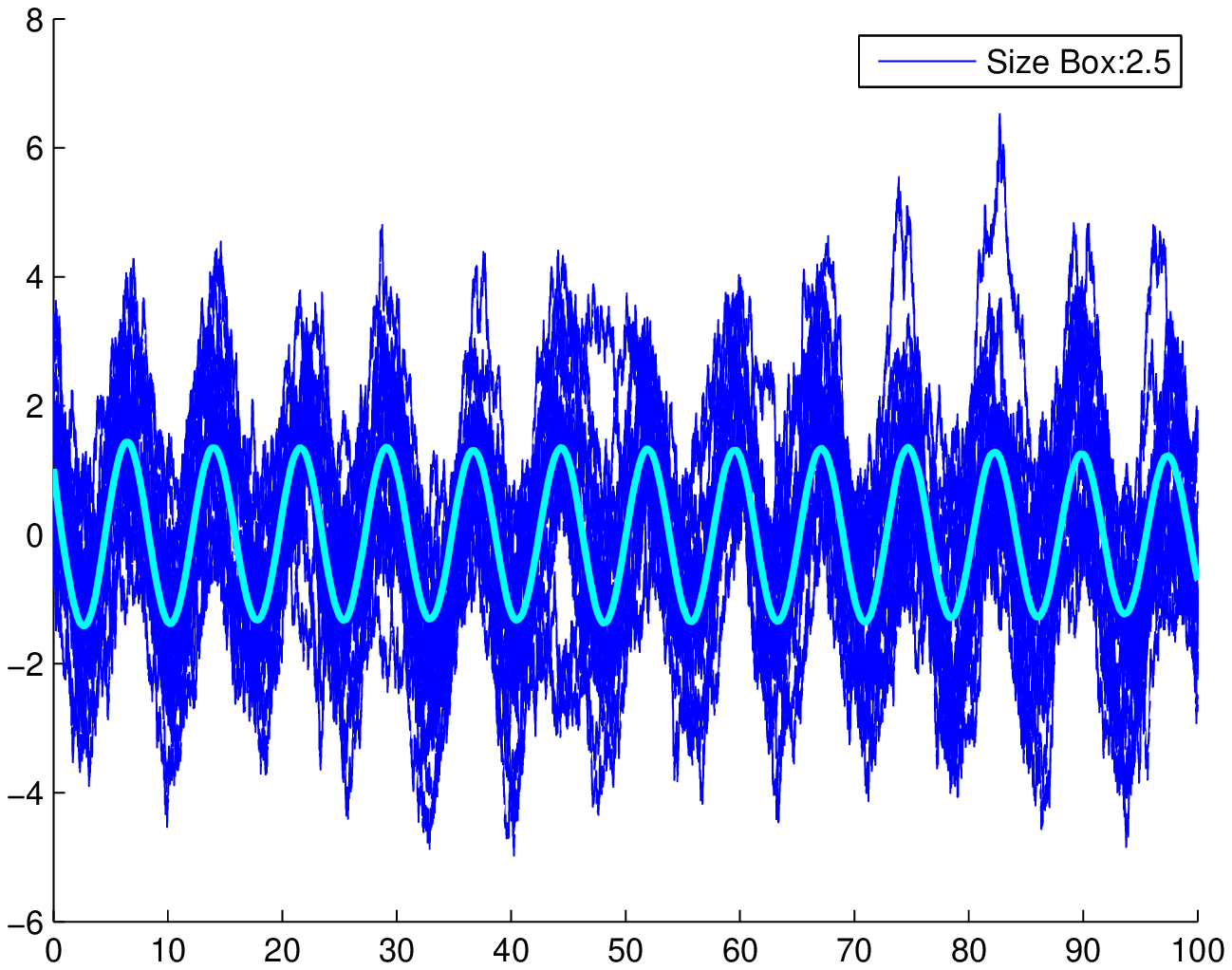}
   \label{fig:osc2a1}
 }
 \centering
 \subfigure[$a=4.5$]{
  \includegraphics[width=0.45\textwidth]{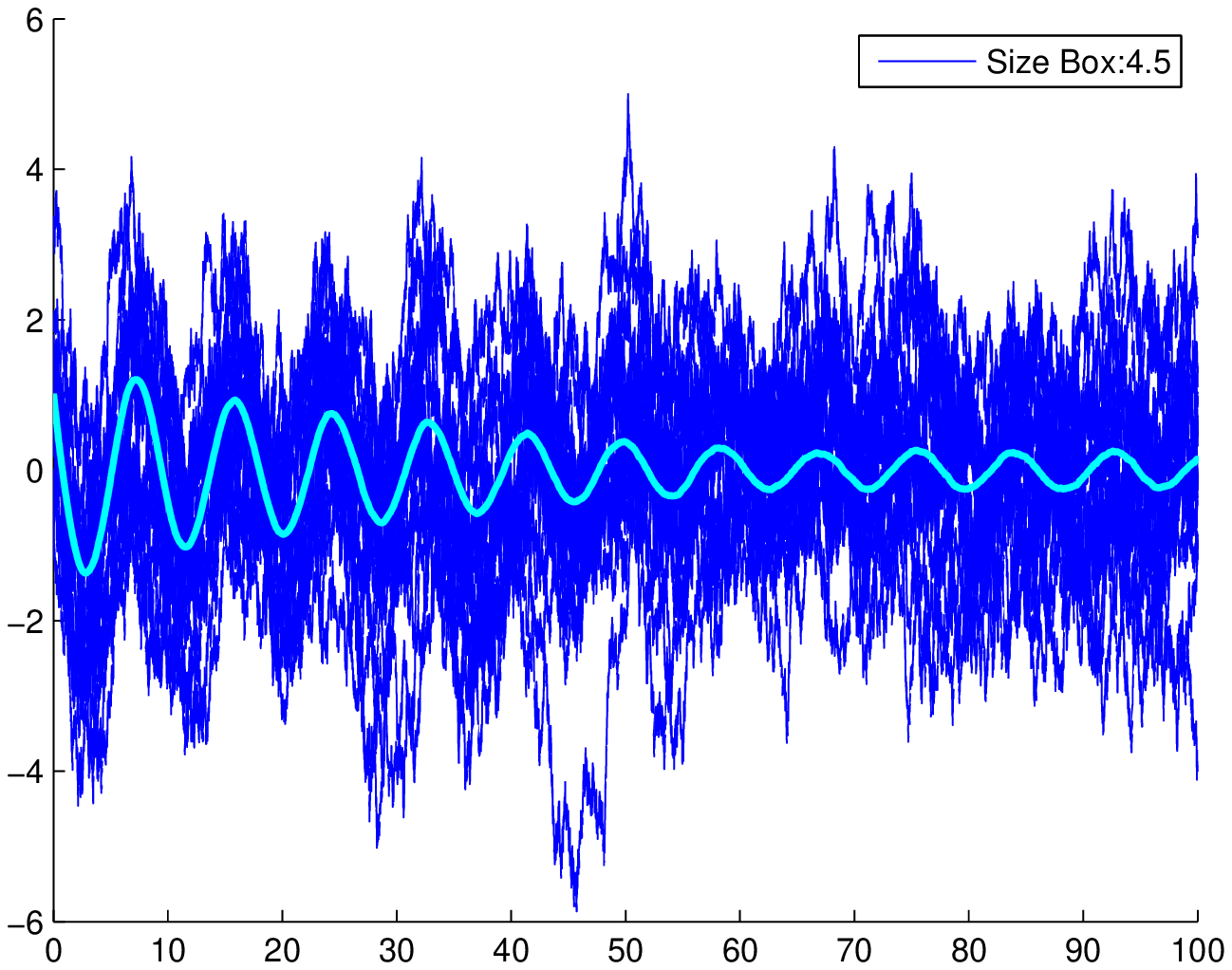}
   \label{fig:osc3a2}
 } 
  \caption{Neurons uniformly distributed in $[0,a]$. Fixed parameters $\theta=3$, $\bar J=-5$, $\lambda=1$. (a) Bifurcation diagram for $\beta=0.1$ in the space $(a,\tau)$: gray zone correspond to oscillatory solutions. For $\tau_s=1.3$: (b-d) Increasing the parameter $a$ (the size of the neural field) induces transition from stationary to periodic and back to stationary. All simulation were made for an Euler explicit method with $N=5000$.}
  \label{fig:sizeVariable}
\end{figure}

In figure~\ref{fig:sizeVariable}, we represent the curve of Hopf bifurcation given by \eqref{eq:Paramao:02} for a fixed value of the parameter $\beta$. This bifurcation diagram separates the parameter space $(a,\tau_s)$ into a region of oscillatory regime and a region of stationary behavior. The typical shape of the Hopf bifurcation curve is a parabola, displaying a unique minimum for a value that we denote by $(a^m,\tau_s^m)$. We denote $\tau_s^0$ the value of the Hopf bifurcation curve for $a=0$ (i.e. fully connected network with deterministic delays $\tau_s$). For $a=0$, the system depends on the delays in the following fashion: for any $\tau_s<\tau_s^0$, the system converges towards stationary behaviors, and for $\tau_s>\tau_s^0$, the system displays periodic behaviors. 

For $\tau_s>\tau_s^0$ fixed, long-range connectivities (corresponding to small values of $\beta$) produce synchronized periodic behaviors that disappear when the network becomes less connected, until a specific value of $\beta$ corresponding to the unique intersection of the Hopf curve with the line of constant $\tau_s$. For $\tau_s^0<\tau_s<\tau_s^m$, the long-range (small $\beta$ or small $a$) and short-range (large $\beta$ or large $a$) connectivity models correspond to stationary behaviors, and for values of the network length $a$ (or range $\beta$) in a specific interval, the system will display synchronized behaviors. Eventually, for $\tau_s<\tau_s^m$, the system only displays stationary solutions whatever the length of the network $a$ or the range $\beta$. 

\subsubsection{The effect of the connectivity factor}

Let us now fix the size of the interval $a>0$. We investigate the effects of $\beta$ and $\tau_s$ on the solution. Equation \eqref{eq:Hopf1} can be written in the form:

\begin{equation}\label{eq:Paramao:08}
\begin{cases}
 \omega^2=-\frac1{\theta^2}+\vert Z(\omega,\beta)\vert^2,\\
 \tau_s=\left(Arg(Z(\omega,\beta))-Arg\left(\frac1{\theta^2}+i\omega\right)+2k\pi\right)\frac{1}\omega
\end{cases}
\end{equation}
with
\[ \displaystyle Z(\omega,\beta)=\frac{2\bar J}{\sqrt{2\pi(1+v^*)}}\frac{1}{a(\beta+i\omega)}\left(1-\frac1{a(\beta+i\omega)}+\frac{e^{-a(\beta+i\omega))}}{a(\beta+i\omega)}\right)\]
We solve this equation by numerically computing the manifold:
\[S_0:=\Big\{(\omega,\beta)\in\R\times\R_+,\mbox{ such that }\omega^2+\frac1{\theta^2}-|Z(\omega,\beta)|^2=0\Big\}\]
from which one can readily compute the delay corresponding to the Hopf bifurcation.
\begin{figure}
\centering
 \subfigure[$a=3$ fixed]{
  \includegraphics[width=0.45\textwidth]{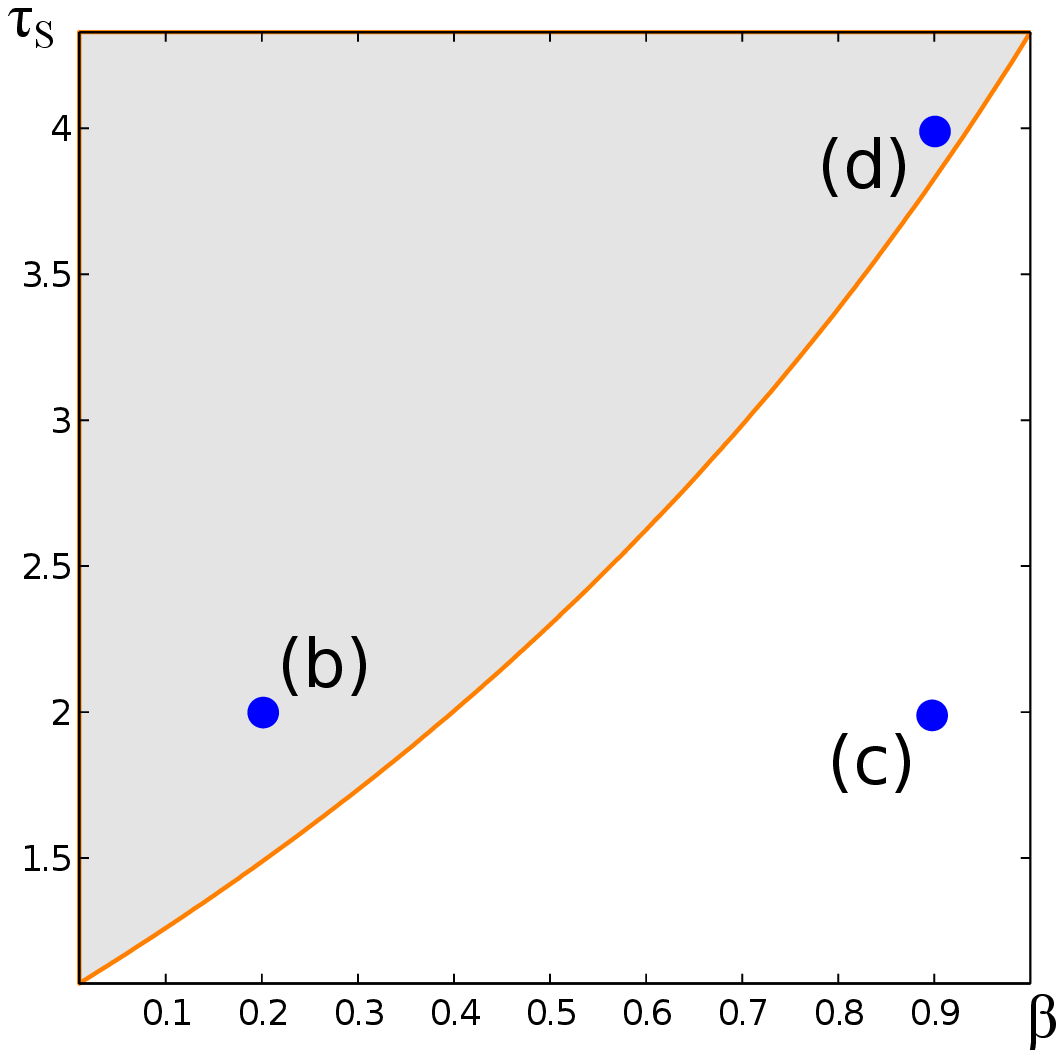}
   \label{fig:hopf3d_2}
 }
 \subfigure[$\beta=0.2,\tau_s=2$]{
  \includegraphics[width=0.45\textwidth]{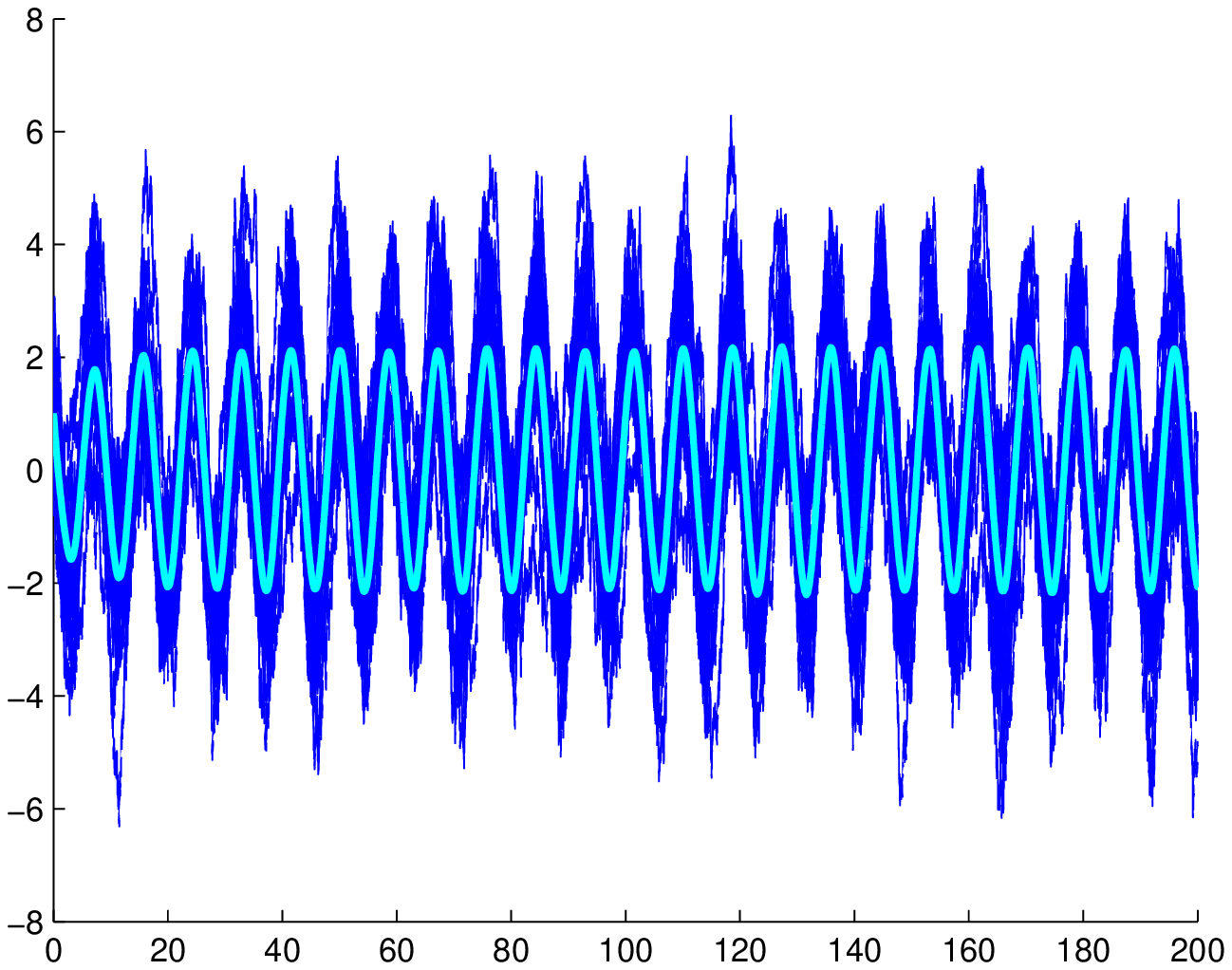}
   \label{fig:hopf2d_2}
 }
 \subfigure[$\beta=0.9,\tau_s=2$]{
  \includegraphics[width=0.45\textwidth]{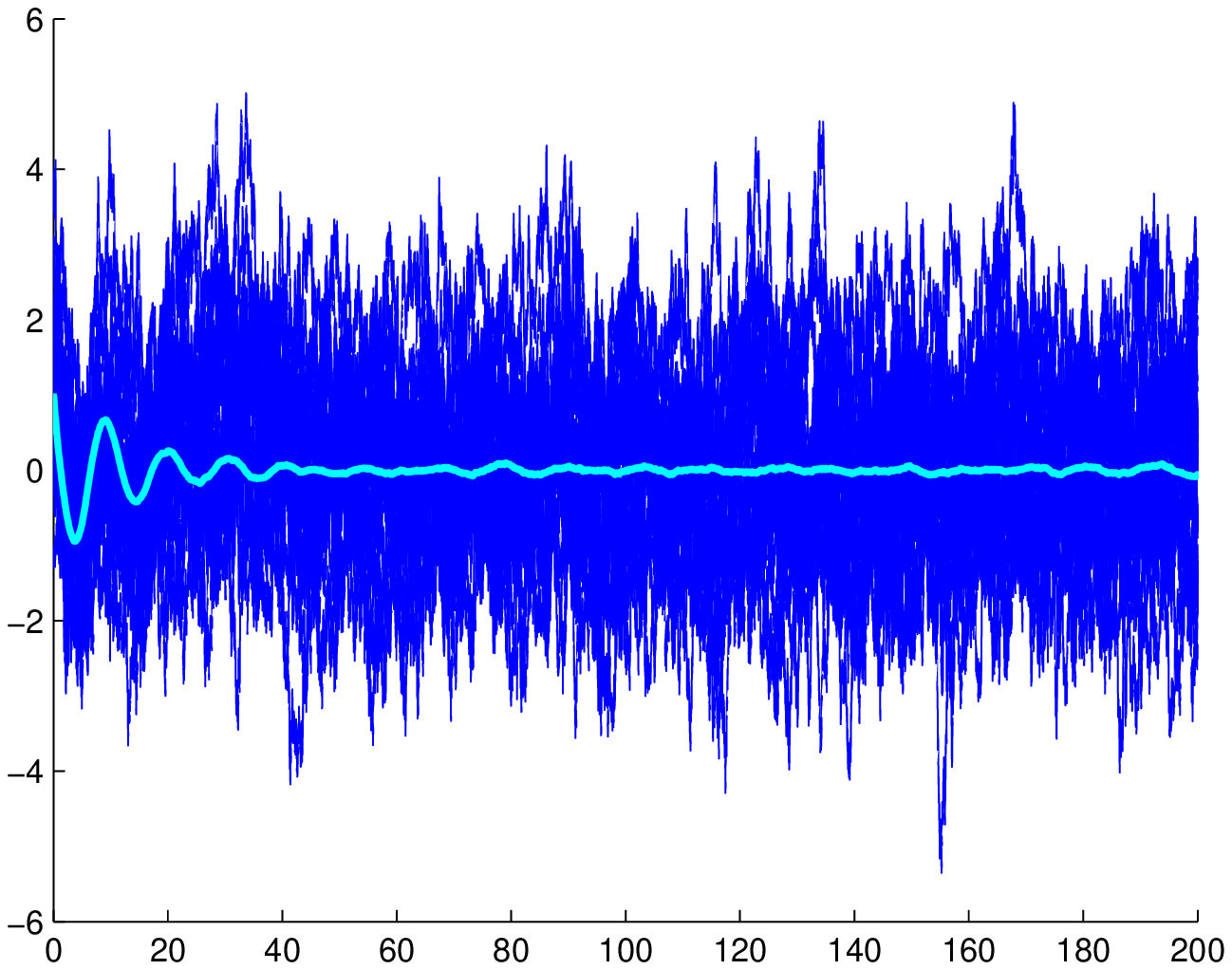}
   \label{fig:osc1}
 }
 \subfigure[$\beta=0.9,\tau_s=4$]{
 \includegraphics[width=0.45\textwidth]{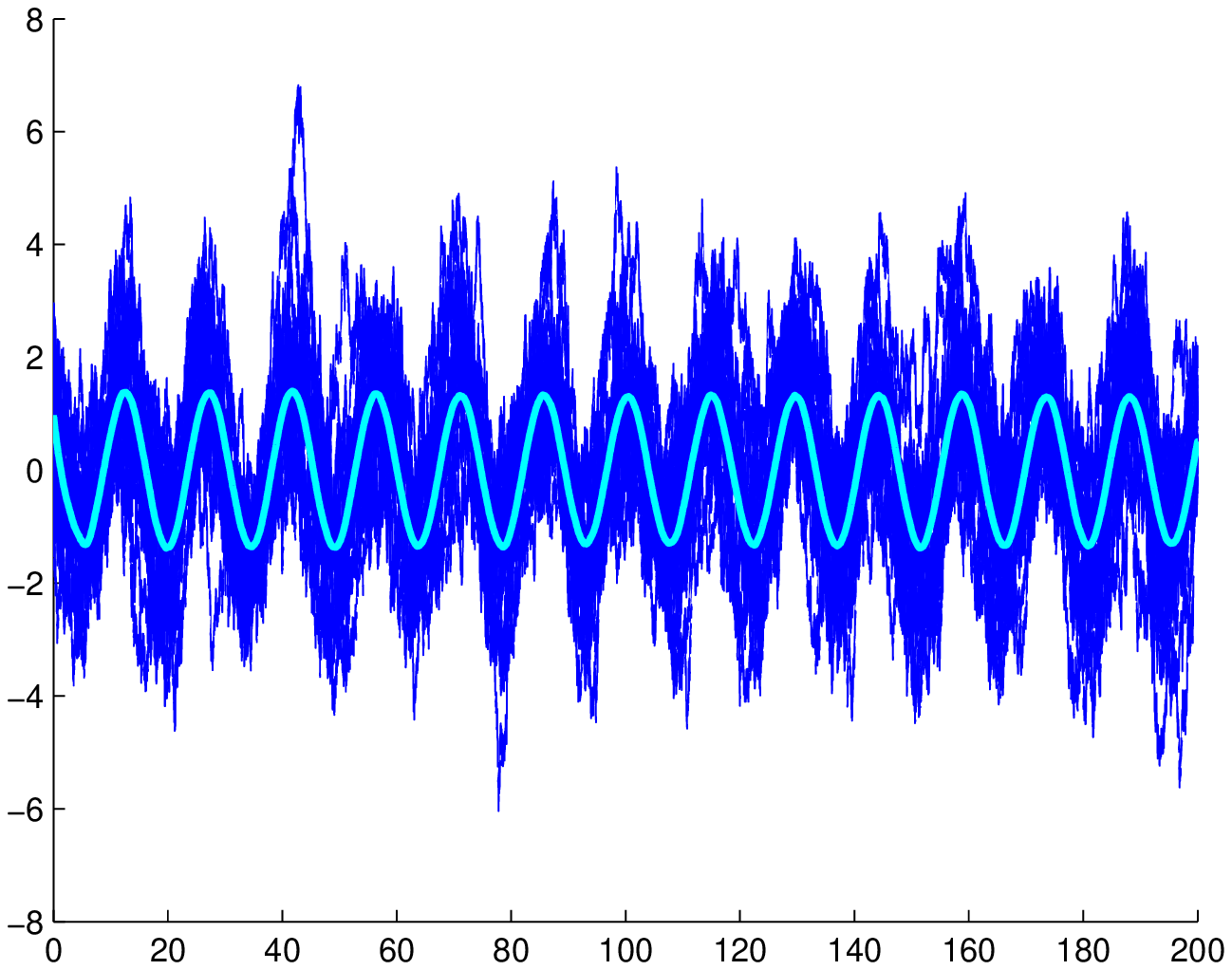}
   \label{fig:osc2}
 }
\caption{Neurons uniformly distributed in $[0,a]$ for different values of $\beta$. Fixed parameters $\theta=1$, $\bar J=-3.5$, $\lambda=0.5$: (a) Hopf bifurcation diagram in the plane $(\beta,\tau_s)$ for $a=3$. (c-e) Starting from a point inside the oscillation zone increasing one of the parameters $\beta$ or $\tau_s$ induces transition to the stationary state. All simulation were made for an Euler explicit method with $N=3500$.}
\end{figure}
Figure \ref{fig:hopf3d_2} show the solution to the system \eqref{eq:Paramao:08} for a fixed value of the spatial extension of the neural field. The curve is relatively different: it now appears to be a monotone non-decreasing map separating oscillatory and stationary behaviors. Qualitatively, the global picture remains unchanged: oscillations vanish as $\beta$ is increased, i.e. as the network is less connected. 

\subsubsection{Discussion}
We therefore observe that the topology of the network strongly impacts the collective behavior of the network. For a fixed value of the connectivity parameter, we have seen that there exists an optimal neural field size for synchronization. At this size, the constant delays necessary to induce oscillations is minimal. In contrast, at fixed values of $a$, we observe that the optimal connectivity level ensuring minimal constant delay to induce oscillations is zero: fully connected networks synchronize more easily. In the cortex, for energetic reasons, full connectivity is not favored, and therefore this indicates optimal cluster sizes for synchronization. 

\section{Proofs}\label{sec:Proofs}

We start by showing the well-posedness of the network system stated in proposition~\ref{prop1}:
\begin{proof}[Proposition~\ref{prop1}]
The proof splits into two main steps: we show {\it a priori estimates} and define a {\it contraction map} that implies existence and unicity for a stopped version of the problem. 
	
	\textit{A priori estimates}
	Let us start by showing that all possible solutions of the system have bounded second moment. It is important to remark that the number of particles of the system is fixed. Let $X^N$ be a solution of \eqref{eq:Network} and $\tau_n$ the first time the process $|X^N_t|$ exceeds the quantity $n$. We look for an upper-bound of the form:
	\begin{equation}\label{eq:gronwals}
	\EE\Big[|X_{t\wedge\tau_n}^{N}|^2\Big]\leq\EE\Big[|X_0(0)|^2\Big]+C\int_0^{t\wedge\tau_n}\EE\Big[1+|X_{s\wedge\tau_n}^{N}|^2\Big]\,ds,
	\end{equation}
	where the positive constant $C$ does not depend on $\tilde X_t = (X_s, s\in[t-\tau,t])$ nor on $n$.

	It is clear that It\^o's formula is valid for $|X^N_{t\wedge\tau_n}|^2$ and that we can study each $i\in\{1,\ldots,N\}$ separately. For all $t>0$:
	\begin{eqnarray*}
	|X_{t\wedge\tau_n}^{i,N}|^2&=&|X_0^i|^2+2\int_0^{t\wedge\tau_n}(X_s^{i,N})^Tg_\alpha(s,X_s^{i,N})\,dW_s^i\\
	&&\qquad+2\int_0^{t\wedge\tau_n}\Big[(X^{i,N}_s)^Tf_\alpha(s,X^{i,N}_s)+\frac12|g_\alpha(s,X_s^{i,N})|^2\\
	&&\qquad\qquad+(X^{i,N}_s)^T\sum_{\gamma=1}^{P}\sum_{p(j)=\gamma}\frac{1}{N_\gamma}\big[b_{\alpha\gamma}(w_{ij},X_s^{i,N},X_{s-\tau_{ij}}^{j,N})\big]\Big]ds,
	\end{eqnarray*}
	The stochastic integral has null expectation and a direct application of (H4) allow us to find upper-bounds for the two first lines of the previous equality. The last term is controlled  using (H3):
	\begin{eqnarray*} 
		&&\int_0^{t\wedge\tau_n}(X^{i,N}_s)^T\sum_{\gamma=1}^{P}\sum_{p(j)=\gamma}\frac{1}{N_\gamma}\big[b_{\alpha\gamma}(w_{ij},X_s^{i,N},X_{s-\tau_{ij}}^{j,N})\big]ds\\
		&&\qquad\qquad\leq\int_0^{t\wedge\tau_n}\sum_{\gamma=1}^{P}\sum_{p(j)=1}\frac{1}{N_\gamma}\Big(\bar K_{\alpha\gamma}(w_{ij})+|X^{i,N}_s|^2\Big)\,ds\\
		&&\qquad\qquad\leq P\int_0^{t\wedge\tau_n}\Big(\bar K+|X^{i,N}_s|^2\Big)\,ds,
	\end{eqnarray*}
	where we have introduced $\bar K:=\max_{(\alpha,\gamma)}\max_{(i,j)}\bar K_{\alpha\gamma}(w_{ij})$. Summing over $i$ yields directly to~\eqref{eq:gronwals}.

	Applying Gronwall'"s lemma we find a uniform upper bound for the second moment of $X_{t\wedge\tau_n}$ for any $t\in[-\tau,T\wedge\tau_n]$. Finally letting $n\rightarrow\infty$ provides that for any realization of the synaptic weights and delays the solutions of~\eqref{eq:Network} have bounded second moment.
	
	\textit{Existence.}
	Let $X^0\in\mathcal M^2(C_\tau)$ such that $X^0|_{[-\tau,0]} \stackrel{\mathcal L}{=}X_0$ a given stochastic process. We introduce the map $\Phi$ given by
	\begin{equation}\nonumber
	\Phi:\left\{\begin{array}{lll}
	 \mathcal{M}(\mathcal{C})&\mapsto&\mathcal{M}(\mathcal{C}) \\
	 X & \mapsto&(Y_t=\{Y_t^{i,N},i=1,\ldots,N\}),\mbox{ with} \\
	 &&\begin{array}{ll}
	 Y_t^{i,N}=&\displaystyle X_0^{i,N}(0)+\int_0^t\Big(f_\alpha(s,X_s^{i,N})\\
	 &+\displaystyle\sum_{\gamma=1}^P\sum_{p(j)=\gamma}\frac1{N_\gamma}b_{\alpha\gamma}(w_{ij},X_s^{i,N},X_{s-\tau_{ij}}^{j,N})\Big)ds\\
	  &+\displaystyle\int_0^tg_\alpha(s,X_s^{i,N})\cdot dW_s^i;\quad t>0\\
	 Y_t=&X_0^{i}(t),\quad t\in[-\tau,0]
	 \end{array}
	\end{array}\right.
	\end{equation}
	and the sequence of processes $(X^k)_{k\geq0}$ on $\mathcal M(\mathcal C)$ given by the induction $X^{k+1}=\Phi(X^k)$. Existence and uniqueness are classically shown through a fixed point argument on the map $\Phi$. 
	
	For compactness of notations, we denote $X^{i,k}_t\in E$ the $i$ component of the vector $X^k_t$. We decompose the difference into elementary terms:
	\begin{eqnarray*}
	X_t^{i,k+1}-X_t^{i,k}&=&\int_0^t\big(f_\alpha(s,X_s^{i,k})-f_\alpha(s,X_s^{i,k-1})\big)ds\\
	&&+\int_0^t\sum_{\gamma=1}^{P(N)}\sum_{p(j)=\gamma}\frac1{N_\gamma}\Big[ b_{\alpha\gamma}\big(w_{ij}, X_s^{i,k},X_{s-\tau_{ij}}^{j,k}\big)\\
	&& \qquad\qquad-b_{\alpha\gamma}\big(w_{ij},X_s^{i,k-1}, X_{s-\tau_{ij}}^{j,k}\big)\Big]\,ds\\
	&&+\int_0^t\sum_{\gamma=1}^{P(N)}\sum_{p(j)=\gamma}\frac1{N_\gamma}\Big[ b_{\alpha\gamma} \big(w_{ij},X_s^{i,k-1},X_{s-\tau_{ij}}^{j,k}\big)\\
	&& \qquad\qquad-b_{\alpha\gamma}\big(w_{ij},X_s^{i,k-1}, X_{s-\tau_{ij}}^{j,k-1}\big)\Big]\,ds\\
	&&+\int_0^t\big(g_\alpha(s,X_s^{i,k})-g_\alpha(s,X_s^{i,k-1})\big)\cdot dW_s^i\\
	&\stackrel{\mbox{{\tiny def}}}{=}& A_t^i+B_t^i+C_t^i+D_t^i,
	\end{eqnarray*}
	where we simply identify each of the four terms $A_t=(A_t^i, i=1,\ldots,N)$, $B_t,$ $C_t,$ and $D_t$ with their corresponding expression. Using Holder's inequality  $$|X_t^{k+1}-X_t^k|^2\leq 4(|A_t|^2+|B_t|^2+|C_t|^2+|D_t|^2),$$ and treat each term separately.
	The first term $A_t$ and the last term $D_t$ are easily controlled using standard techniques (Cauchy-Schwarz inequality and Burkholder-Davis-Gundy theorem) and (H1). In $B_t$ follows
	\begin{eqnarray*}
	\sum_{i=1}^N\left|\sum_{\gamma=1}^P\sum_{p(j)=\gamma}\int_0^t\frac1{N_\gamma}\left[b_{\alpha\gamma}(w_{ij},X_s^{i,k},X_{s-\tau_{ij}}^{j,k})-b_{\alpha\gamma}(w_{ij},X_s^{i,k-1},X_{s-\tau_{ij}}^{j,k})\right]ds\right|^2\\
	 \leq\sum_{i=1}^NPt\sum_{\gamma=1}^P\sum_{p(j)=\gamma}\int_0^t\left|b_{\alpha\gamma}(w_{ij},X_s^{i,k},X_{s-\tau_{ij}}^{j,k})-b_{\alpha\gamma}(w_{ij},X_s^{i,k-1},X_{s-\tau_{ij}}^{j,k})\right|^2ds\\
	 \leq TP^2L^2N\int_0^t\big|X_s^{k}-X_s^{k-1}\big|^2ds,
	\end{eqnarray*} where $L:=\max_{(\alpha,\gamma)}L_{\alpha\gamma}$.
and similarly for $C_t$.

The conclusion is easy, at this point we have:
	\begin{eqnarray}\label{eq:InQuad}
	\EE\big[\sup_{-\tau<s<t}\big|X_s^{k+1}-X_s^{k}\big|^2\big]&\leq& C\int_0^t\EE\big[\sup_{-\tau\leq u\leq s}|X_u^{k}-X_u^{k-1}|^2\big]ds, 
	\end{eqnarray}
where $C>0$ depends on $T,K,L$ and $P$. Calling $$M^k_t\stackrel{\mbox{{\tiny def}}}{=}\EE\big[\sup_{-\tau\leq s\leq t}|X_s^{k}-X_s^{k-1}|^2\big],$$ {\it a priori bounds} ensures that $M^0_T<\infty$ and the recursive inequality holds 
\begin{eqnarray*}
 M_t^k&\leq& C^k\int_0^t\int_0^{s_1}\ldots\int_0^{s_{k-1}}M_{s_k}^0ds_1\ldots ds_k\leq C^k\frac{t^k}{k!}M_T^0,
 \end{eqnarray*}
From the last inequality we get that $$\sum_{n=1}^\infty\EE\Big[\sup_{-\tau\leq s\leq t}\big|X_s^{n+1}-X_s^{n}\big|^2\Big]<\infty,$$ which implies in particularly the almost sure convergence of $$X_t^0+\sum_{k=0}^n(X_t^{k+1}-X_t^{k})=X_t^n,$$ on $[-\tau,T]$. The limit defined $\bar X_t$ is trivially a fixed point of $\Phi$ and by consequence solution to networks equations~\eqref{eq:Network}.

	\textit{Uniqueness.}
	Starting  with two solutions of the network equations \eqref{eq:Network} with exactly the same initial condition one can remake the argument used to find \eqref{eq:InQuad} and then the uniqueness follows directly from Gronwall's lemma.\hfill$\square$
	
\end{proof}

The proof well-posedness of mean field equation~\eqref{eq:MeanField2} (Theorem~\ref{theo1}) is very similar:

\begin{proof}[Theorem~\ref{theo1}]
	It might seem that averaging over the delays and weights could add some new technical difficulties to the upper-bounds for the second moment but thanks to $(H3)$ similar estimates hold.
	
	To illustrate how to deal with our random network framework, let $X$ be a solution of the mean-field equations and once again $\tau_n$ the first time that the process $|X_t|$ exceeds the quantity $n$. 
	Applying It\^o's formula to $|X_{t\wedge\tau_n}|^2$ we now find
	\begin{eqnarray*}
	|X_{t\wedge\tau_n}^\alpha|^2&=&|\zeta_0^\alpha|^2+2\int_0^{t\wedge\tau_n}\Big[(X^\alpha_t)^Tf_\alpha(s,X^\alpha_s)+\frac12|g_\alpha(s,X_s^\alpha)|^2\\
	&&+(X_s^\alpha)^T\sum_{\gamma=1}^P\int_{-\tau}^0\int_{\R}\Big[\EE_{\bar Y}\big[b_{\alpha\gamma}(w,X_s^\alpha,\bar Y_{s+u}^\gamma)\big]d\Lambda_{\alpha\gamma}(u,w)\Big]ds\\
	&&+2\int_0^{t\wedge\tau_n}(X_s^\alpha)^Tg_\alpha(s,X_s^\alpha)\,dW_t^\alpha,
	\end{eqnarray*}
	the only interesting term is the one in the second line, using triangular inequality and $(H3)$ we get
\begin{eqnarray*}
&&(X_s^\alpha)^T\sum_{\gamma=1}^P\int_{-\tau}^0\int_{\R}\Big[\EE_{\bar Y}\big[b_{\alpha\gamma}(w,X_s^\alpha,\bar Y_{s+u}^\gamma)\big]d\Lambda_{\alpha\gamma}(u,w)\Big]ds\\
&&\qquad\leq P|X_s^\alpha|^2+\sum_{\gamma=1}^P\int_{-\tau}^0\int_{\R} \bar{K}_{\alpha\gamma}(w)d\Lambda_{\alpha\gamma}(u,w)\leq C(\bar k+|X_s^\alpha|^2).
\end{eqnarray*}
	Equipped with this estimate, the proof is identical to that of the related property in proposition~\ref{prop1}, i.e., define a contraction mapping which gives the existence and uniqueness of solutions. \hfill$\square$

\end{proof}

The two following proofs deal with the propagation of chaos property, we first demonstrate Theorem~\ref{theo2} which states the convergence properties in a quenched sense in the translation invariant case, and we finally explain how to adapt this proof to the general case Theorem~\ref{theo3}, i.e., how to deal with the additional difficulty of averaging over all possibles positions of neurons in each population.

\begin{proof}[Theorem~\ref{theo2}] The idea extends standard arguments for propagation of chaos and mean-field limits by considering random correlated coupling and delays. The argument remains to control the difference between the two processes as $N$ goes to infinity. Decomposing the difference in 5 simpler terms we find:
\begin{eqnarray*}
 X^{i,N}_t-\bar X^i_t &=& \int_0^t\big(f_\alpha(s,X^{i,N}_s)-f_\alpha(s,\bar X^i_s)\big)\,ds\\
 &&+\int_0^t\big(g_\alpha(s,X^{i,N}_s)-g_\alpha(s,\bar X^i_s)\big)\cdot dW_s^i\\
 &&+\sum_{\gamma=1}^P\int_0^t\sum_{p(j)=\gamma} \big[b_{\alpha\gamma}(w_{ij},X_s^{i,N},X^{j,N}_{s-\tau_{ij}})-b_{\alpha\gamma}(w_{ij},\bar X_s^{i},X^{j,N}_{s-\tau_{ij}})\big]\frac{ds}{N_\gamma}\\
 &&+\sum_{\gamma=1}^P\int_0^t\sum_{p(j)=\gamma} \big[b_{\alpha\gamma}(w_{ij},\bar X_s^{i},X^{j,N}_{s-\tau_{ij}})-b_{\alpha\gamma}(w_{ij},\bar X_s^{i},\bar X^{j}_{s-\tau_{ij}})\big]\frac{ds}{N_\gamma}\\
 &&+\sum_{\gamma=1}^P\int_0^t\Big(\frac1{N_\gamma}\sum_{p(j)=\gamma} b_{\alpha\gamma}(w_{ij},\bar X_s^{i},\bar X^{j}_{s-\tau_{ij}})\\
&& \qquad\qquad-\int_{-\tau}^0\int_\R\EE_Z\big[b_{\alpha\gamma}(w,\bar X^i_s,Z^\gamma_{s+u})\big]d\Lambda_{\alpha\gamma}(u,w)\Big)ds\\
 &:=&A_t(N)+B_t(N)+C_t(N)+D_t(N)+E_t(N).
\end{eqnarray*}

We are interested in the behavior of $\E[\EE(\sup_{-\tau\leq s\leq T}|X^{i,N}_s-\bar X^i_s|^2)]$ as $N\rightarrow\infty$. Under the same ideas used in Proposition~\ref{prop1} and in Theorem~\ref{theo1}, we find:
\begin{eqnarray*}
 \EE[\sup_{0\leq s\leq t}|A_s(N)|^2]&\leq& K'^2T\int_0^{t}\EE[\sup_{-\tau\leq u\leq s}|X^{i,N}_u-\bar X^i_u|^2]\,ds\\
  \EE[\sup_{0\leq s\leq t}|B_s(N)|^2]&\leq& 4K'^2\int_0^{t}\EE[\sup_{-\tau\leq u\leq s}|X^{i,N}_u-\bar X^i_u|^2]\,ds,\\ \EE[\sup_{0\leq s\leq t}|C_s(N)|^2]&\leq& TL^2P^2\int_0^{t}\EE[\sup_{-\tau\leq u\leq s}|X^{i,N}_u-\bar X^i_u|^2]\,ds\\
\EE[\sup_{0\leq s\leq t}|D_s(N)^2]&\leq& TL^2P^2\int_0^{t}\max_{k=1,\ldots,N}\EE[\sup_{-\tau\leq u\leq s}|X^{k,N}_u-\bar X^k_u|^2]\,ds,
\end{eqnarray*} where $L$ is the maximum value of constants $L_{\alpha\gamma}$ (finite number of populations) and we precise that the 4 in the $B_t(N)$ upper-bound is found using the Burkholder-David-Gundy Inequality.

For the last term $E_t(N)$ we start by applying the Cauchy-Schwartz and the triangular inequality:
\begin{eqnarray*}
\E[\EE[\sup_{0\leq s\leq t}|E_s(N)|^2]]&&\leq TP\sum_{\gamma=1}^P\int_0^t\E\Big[\EE\Big[\Big|\frac1{N_\gamma}\sum_{p(j)=\gamma} \Big(b_{\alpha\gamma}(w_{ij},\bar X^i_s,\bar X^j_{s-\tau_{ij}})\\
&& \qquad\qquad-\int_{-\tau}^0\int_{\R}\EE_Z[b_{\alpha\gamma}(w,\bar X^i_s,Z^\gamma_{s+u})]d\Lambda_{\alpha\gamma}(u,w)\Big)\Big|^2\Big]\Big]\,ds,
\end{eqnarray*}
moreover,
\begin{eqnarray*}
&&\E\Big[\EE\Big[\Big|\frac1{N_\gamma}\sum_{p(j)=\gamma} \Big(b_{\alpha\gamma}(w_{ij},\bar X^i_s,\bar X^j_{s-\tau_{ij}})\\
&&\qquad-\int_{-\tau}^0\int_{\R}\EE_Z[b_{\alpha\gamma}(w,\bar X^i_s,Z^\gamma_{s+u})]d\Lambda_{\alpha\gamma}(u,w)\Big)\Big|^2\Big]\Big]\\
&&=\frac1{N_\gamma^2}\sum_{p(j)=\gamma}\sum_{p(l)=\gamma}\E\Big[\EE\Big[\Big(b_{\alpha\gamma}(w_{ij},\bar X^i_s,\bar X^j_{s-\tau_{ij}})-\EE_{Z,(\tilde\tau,\tilde w)_{\alpha\gamma}}[b_{\alpha\gamma}(\tilde w_{\alpha\gamma},\bar X^i_s,Z^\gamma_{s-\tilde\tau_{\alpha\gamma}})]\Big)^T\cdot\\
&&\qquad\qquad\qquad\Big(b_{\alpha\gamma}(w_{il},\bar X^i_s,\bar X^l_{s-\tau_{il}})-\EE_{Z,(\tilde\tau,\tilde w)_{\alpha\gamma}}[b_{\alpha\gamma}(\tilde w_{\alpha\gamma},\bar X^i_s,Z^\gamma_{s-\tilde\tau_{\alpha\gamma}})]\Big)\Big]\Big]
\end{eqnarray*}
In the above expression, $(\tilde\tau,\tilde w)_{\alpha\gamma}$ denotes a random variable with law $\Lambda_{\alpha\gamma}$ independent of the sequence of delays, weights and Brownian motions. We remark that $\int_{-\tau}^0\int_\R\EE_Z[b_{\alpha\gamma}(w,\bar X^i_s,Z^\gamma_{s+u})]d\Lambda_{\alpha\gamma}(u,w)$ is exactly the expectation of $b_{\alpha\gamma}(w_{ij},\bar X^i_s,\bar X^j_{s-\tau_{ij}})$ under the law of $\bar X^j$ and of the pair delays-weights. 

Therefore in the case $j\neq l$, the term in the summation vanishes, and in the opposite case $j=l$ we use the triangular inequality to see that
\begin{eqnarray*}
&&\E\Big[\EE\Big[\big|b_{\alpha\gamma}(w_{ij},\bar X^i_s,\bar X^j_{s-\tau_{ij}})-\EE_{Z,(\tilde\tau,\tilde w)_{\alpha\gamma}}[b_{\alpha\gamma}(\tilde w_{\alpha\gamma},\bar X^i_s,Z^\gamma_{s-\tilde\tau_{\alpha\gamma}})]\big|^2\Big]\Big]\\
&&\leq2\,\E\Big[\EE\Big[\big|b_{\alpha\gamma}(w_{ij},\bar X^i_s,\bar X^j_{s-\tau_{ij}})\big|^2+\big|\EE_{Z,(\tilde\tau,\tilde w)_{\alpha\gamma}}[b_{\alpha\gamma}(\tilde w_{\alpha\gamma},\bar X^i_s,Z^\gamma_{s-\tilde\tau_{\alpha\gamma}})]\big|^2\Big]\Big]\\
&&\leq2\,\E\big[\EE\big[\bar K(w_{ij})+\bar k\big]\big]\leq 4\bar k.
\end{eqnarray*}

This implies that number of non-null terms in the sum is proportional to $N_\gamma$ and all of them are bounded by the same quantity. Thus
\begin{equation}\nonumber
\E\big[\EE\big[\sup_{0\leq s\leq t}|E_s(N)|^2\big]\big]\leq \bar Ck\sum_{\gamma=1}^P\frac{1}{N_\gamma}\leq\frac{C\bar k P}{\min_\gamma(N_\gamma)}.
\end{equation}

Assembling all the estimates, using that on $[-\tau,0]$ both $X^{i,N}_t$ and $\bar X^i_t$ are equal and denoting by $C$ any generic constant that does not depend on $N$ we find
\begin{multline}\nonumber
 \max_{i=1,\ldots,N}\E\big[\EE\big[\sup_{-\tau\leq s\leq t}|X_s^{i,N}-\bar X^i_s|^2\big]\big]\\ \leq C\int_0^t\max_{k=1,\ldots,N}\E\big[\EE\big[\sup_{-\tau\leq u\leq s}|X_u^{k,N}-\bar X^k_u|^2\big]\big]ds+\frac{C}{\min_\gamma(N_\gamma)},
\end{multline}
by Gronwall's inequality:
\begin{equation}
 \max_{i=1,\ldots,N}\E\big[\EE\big[\sup_{-\tau\leq s\leq t}|X_s^{i,N}-\bar X^i_s|^2\big]\big]\leq \frac{C e^{C T}}{\min_\gamma(N_\gamma)},
\nonumber
\end{equation}
which tends to zeros as $N$ goes to infinity by (H0).

As a side result, the almost sure convergence towards the coupled process implies the convergence in law of $(X^{i,N}_{t},-\tau\leq t\leq T)$ towards $(\bar X^{\alpha}_{t},-\tau\leq t\leq T)$.

\hfill$\square$
\end{proof}

From the last inequality we have easily the propagation of chaos property. Fixing a finite set of neurons $(i_1,\ldots,i_l)\in\mathbb N$, then if $f_{\alpha}$ and $g_{\alpha}$ are globally Lipschitz continuous, we have:
\begin{equation}
 \max_{i_1,\cdots, i_l\in\{1,\ldots,N\}^l}\E\big[\EE\big[\sup_{-\tau\leq s\leq t}|(X_s^{i_1,N},\ldots,X_s^{i_l,N})-(\bar X^{i_1}_s,\ldots,\bar X^{i_l,N}_s)|^2\big]\big]\leq \frac{lC e^{C T}}{\min_\gamma(N_\gamma)},
\nonumber
\end{equation}
hence $$(X_s^{i_1,N},\ldots,X_s^{i_l,N},-\tau\leq s\leq T)\xrightarrow{\mathcal{L}}(\bar X^{i_1}_s,\ldots,\bar X^{i_l,N}_s,-\tau\leq s\leq T),$$ and truncation argument allows to conclude on the convergence in the locally Lipschitz case. This implies that the vector $(X_s^{i_1,N},\ldots,X_s^{i_l,N},-\tau\leq s\leq T)$ converges in law towards $m^{i_1}\otimes\ldots\otimes m^{i_l}$, readily implying propagation of chaos.

\begin{proof}[Theorem~\ref{theo3}] The proof uses essentially the same arguments as that of theorem~\ref{theo2}. Here, we control the difference between $\mathcal E_i[X^{i,N}_t]$ and $\bar X^i_t$ in the quadratic norm $\|Z\|^2:=\EE[\sup_{-\tau\leq t\leq T}|Z_s|^2]$. The assumption on $b$ allow us to separate the distance into only 4 terms similarly to the quenched case. Most terms are handled in a similar fashion, the only difference being the presence of a additional expectation $\E_i$. The main difference is to deal with the term corresponding to $E_t(N)$, which now reads:
	\begin{eqnarray*}
	 &&\E\big[\EE\big[\sup_{0\leq s\leq t}|E_s(N)'|^2\big]\big]=\\
	 &&\qquad\leq TP\sum_{\gamma=1}^P\int_0^t\E\Big[\EE\Big[\Big|\frac1{N_\gamma}\sum_{p(j)=\gamma}\mathcal E_i[\ell_{\alpha\gamma}(w_{ij},\bar X^j_{s-\tau_{ij}})]\\
	 &&\qquad\qquad\qquad-\int_{-\tau}^0\int_\R\EE_Z[\ell_{\alpha\gamma}(w,Z^\gamma_{s+u})]d\Lambda_{\alpha\gamma}(u,w)\,dr_i\Big|^2ds\Big]\Big],
	\end{eqnarray*}
	Again,$$\mathcal E\Big[\EE\Big[\mathcal E_i[\ell_{\alpha\gamma}(w_{ij},\bar X^j_{s-\tau_{ij}})]\Big]\Big]=\int_{-\tau}^0\int_\R\EE_Z[\ell_{\alpha\gamma}(w,Z^\gamma_{s+u})]d\Lambda_{\alpha\gamma}(u,w)$$ we develop in the same way that Theorem~\ref{theo2}. The key point is that it suffices to find an upper-bound uniformly in the disorder of the system which is trivially found using (H3), i.e.,
	\begin{eqnarray*}
&&\E\Big[\EE\Big[\big|\E_i[\ell_{\alpha\gamma}(w_{ij},\bar X^j_{s-\tau_{ij}})]-\EE_{Z,(\tilde\tau,\tilde w)_{\alpha\gamma}}[\ell_{\alpha\gamma}( w_{\alpha\gamma},Z^\gamma_{s-\tilde\tau_{\alpha\gamma}})]\big|^2\Big]\Big]\leq 2\bar k,
\end{eqnarray*}
and we conclude using (H0).\hfill$\square$
\end{proof} 

\section{Discussion}

In this paper, motivated by the structure of interconnection matrix and interactions of neuronal networks of the brain, we analyzed the mean-field limits and dynamics of networks on some random graphs with delays correlated to the synaptic weights. Extending coupling methods to these models, we showed quenched and averaged propagation of chaos, and convergence towards a complex mean-field equation involving distributed delays and averaging with respect to the law of the connectivity. This limit equation is relatively complex in general models, however, they massively simplify for the classical firing-rate model, in which case solutions are exactly reduced to a system of distributed delays integro-differential equations, from which one can infer, using bifurcation theory, the role of random connectivities and delays. This technique led us to demonstrate that typical size of the neuronal area, as well as typical length scale of connectivity, induced or broke synchronization of the neurons. In detail, we showed that depending on the connectivity of the network and the averaged delays the network can either present stationary or a synchronized periodic behavior. In this sense, using a small-world type of model for the value of the weights, we were able to prove that the architecture of the system also plays a role in the dynamics. We also showed that the macroscopic behavior depends on the size of the neural field considered and, more important, on the connectivity of the system measured as the amount of connections over the total possible ones.

\subsection{Relationship with pathological rhythmic brain activity}

Synchronized states are ubiquitous and serve essential function in brain such as memory or attention~\cite{buszaki:06}. Impairments of synchronization levels often relate to severe pathological effects such as epilepsy (too much synchronization) or Parkinson's disease (too little synchronization)~\cite{schnitzler-gross:05}. Troubles in oscillatory patterns have also been related to connectivity levels in epilepsy. In detail, the emergence of seizures and abnormal synchronization was hypothesized to be related to an increased functional connectivity, or more recently to the appearance of an increased number of synaptic buttons between cells. The former phenomenon has been reported in various epileptic situations (see e.g. \cite{bettus-chauvel-etal:08}), and the latter was mainly evidenced in hippocampal epilepsy, and is generally referred to as \emph{neosynaptogenesis}, or \emph{sprouting}, see e.g. \cite{babb-pretorius-etal:89,munoz-mendez-etal:07,noebels:96}. Our models provides an elementary account for the fact that indeed, increased connectivity levels (corresponding to small values of $\beta$) tend to favor synchronization for most values of $\tau_s$. The model even makes a prediction about some possible parameter regions in which this synchronization may only arise in a particular intermediate interval of connectivity levels $\beta$. Disorder also seems to intervene in the emergence of abnormally synchronized oscillations, as evidenced for instance by Aradi and Soltesz~\cite{aradi-soltesz:02} who showed that even if average levels of connectivity in rats subjects to febrile epileptic seizures were similar to those of a control population, variance in the connectivities were increased. Our models incorporate the law of the synaptic weights, and therefore all for testing this hypothesis, as well as a number of variations around these models, in a rigorous manner.

\subsection{Cluster size and synchronization in primary visual area}
The structure of the primary visual areas are very diverse across species. These areas are composed of cells sensitive to the orientation of visual stimuli. In primates, neurons gather into columns as a function of the orientation they are selective to, and these columns organize spatially creating continuous patterns of a specific anatomical size (see e.g. \cite{bosking-zhang-etal:97}). In contrast, rodents present no specific organization of neurons selective to the same orientation (salt-and-pepper organization, see~\cite{ohki-chung-etal:05}). The reason why these architectures are very different across mammals is still poorly understood, and one of the possibles explanations proposed is related to the size of V1: the model tends to show that it is harder to ensure collective synchronization at the level of large cortical areas than locally, phenomenon probably due to the fact that naturally, connectivities are local. This is precisely one of the results of our analysis. In our model, the parameter $a$ characterizes the size of one cortical column, and the results of the analysis of the model show that increasing the size of a column $a$ induces transitions from synchronized regimes to stationary regimes, reducing the collective response of neurons. 

\subsection{Macroscopic vs Mesoscopic models}
The question of which is the proper scale adapted to describe a phenomenon is central in computational neuroscience. Of course, it is tempting to propose large-scale macroscopic models made of homogeneous neuronal populations, as neuronal networks tend to present a columnar organization made of a large number of strongly connected neurons. Most models use implicitly this kind of structure through neural mass models~\cite{wilson-cowan:72,jansen-rit:95}. Another common approximation is the neural field model (see~\cite{bressloff:12} for a recent review) that describes the cortical activity through integro-differential delayed equations, which could be related to a particular limit of neuronal networks with local homogeneity properties as shown in~\cite{touboulNeuralfields:11}. 

The model analyzed sits at an intermediate scale at which homogeneity of connectivity is only true (i) locally an (ii) in a statistical sense. Though these local variations, the model studied in first part of section~\ref{sec:MainResults}, termed \emph{macroscopic}, describes the neural network at a macroscopic scale with a single equation describing the averaged or quenched behavior of one cell in the network. Appendix~\ref{append:InfiniteDimensions} shows that the result persists when considering asymptotically a continuum of neural populations, yielding the \emph{mesoscopic} model. Let us now compare our models to usual neural mass (NM) or neural fields (NF). These latter models are given by the equations (in which $\Phi$ is a sigmoid transform):
\[\dot{u}_{\alpha}(t)=-\frac{u_{\alpha}}{\theta_{\alpha}}(t)+\sum_{\beta=1}^P \bar{J}_{\alpha\beta} \Phi(u_{\beta}(t-\tau_{\alpha\beta}))\]
for finite-populations networks (model NM), and in spatial continuous settings (NF) with a single layer:
\[\partial_t {u}(r,t)=-\frac{u(r,t)}{\theta}+\int_{\Gamma} \bar{J}(r,r') \Phi(u(r',t-\tau(r,r')))\,dr'.\]
These two equations are very close from the mean equations we obtained in our mean-field limit. Disregarding stochastic inputs, the macroscopic (mesoscopic) model is an homogenized version of an heterogeneous neural mass (resp, neural field) model. Disregarding the effect of stochastic noise, our macroscopic model therefore tends to correspond to spatially homogeneous solutions of the neural field equations for translation invariant neural fields. Indeed, assuming $r\in \mathbb{S}_a$ the 1-dimensional torus of length $a$, i.e. the periodic interval $[0,a]$, $J(r,r')=J(r-r')$ and $\tau(r,r')=\tau_s+\vert r-r'\vert$, spatially homogeneous solutions are functions of time only, satisfying the equations:
\[\dot {\bar{u}}(t)=-\frac{\bar{u}(t)}{\theta}+\int_{0}^a \bar{J}(\zeta) \Phi(\bar{u}(t-\tau_s - \zeta)))\,d\zeta\]
(which does not depend on $r$). 
Our model yields an equation on the mean of the process that corresponds to:
\[\dot {\mu}(t)=-\frac{\mu(t)}{\theta}+\int_{0}^a \bar{J} \beta(\zeta) f(\mu(t-\tau_s - \zeta),v(t-\tau_s - \zeta)))\,d\zeta.\]
Therefore, with an appropriate choice of parameters and function, the mean-field macroscopic model represents spatially homogeneous solutions of the Wilson-Cowan neural field equations. The present approach provides a microscopic interpretation of these equations, and the model provides therefore a suitable framework to investigate random individual phenomena arising in large neuronal areas, observed at scales that do not resolve fine structure of the brain, such as the electro-encephalogram method used in epilepsy monitoring. 

\subsection{Perspectives}
The course of our developments lead us to cast aside the assumption of full connectivity or exchangeability between neurons. Incidentally, this work therefore shows that the notion of exchangeability, widely use in large stochastic particle systems, can be significantly weakened, in favor of statistical equivalent, and more structured global exchangeability properties such as the translation invariance. This opens the way to develop a these ideas towards invariant architectures under the action of specific groups of transformation. This constitute an active research that we are currently developing. This method also has a number of possible implications in neuroscience and in complex systems more generally, and may help understanding the dynamics of large neural networks. Enriching this model considering different populations in the applications section is a straightforward extension of the manuscript, and analyzing those results would allow going even deeper in the analysis of neuronal networks and macroscopic synchronization of them as an effect of random pairs delays and synaptic weights. Considering different kind of architectures is also a possible path to follow and could bring new relationships with the specific cortical functions. A deep question is whether one can obtain information on the microscopic configurations related to the macroscopic regimes observed. This motivates to develop the analysis of the presence of structured activity (localized bumps, traveling waves, traveling pulses) and their probability of appearance as a function of disorder, noise and the parameters of the system. This is an exciting question well worth investigating. One limitation of the qualitative analysis provided here is that the moment reduction is rigorously exact only in very specific models where solutions are Gaussian. Such models do not reproduce the excitability properties of the cells. Extending this analysis to excitable systems, i.e. analyzing equation~\eqref{eq:MeanField1} with nonlinear dynamics and nonlinear interactions, is a deep and challenging mathematical question in the domain of stochastic processes and functional analysis.

\appendix
\section{Randomly connected neural fields}\label{append:InfiniteDimensions}\label{Append:NeuralField}
We now extend the above results to the mesoscopic case of spatially-extended neural fields with random correlated connectivity weights and delays. In this case, following~\cite{touboulNeuralfields:11}, we consider that the number of populations in a network of size $N$ is $P(N)$, and this quantity diverges when $N$ tends to infinity covering, in the limit $N\to\infty$, a piece of cortical tissue $\Gamma$ which compact set of $\R^\delta$ (generally $\delta=1,2$). In this interpretation, a population index represents the location $r_{\alpha}\in\Gamma$ of a microcolumn on the neural field, which are assumed to be independent random variables with distribution $\lambda$ on $\Gamma$. For the sake of simplicity and consistency with other works about neural fields, we include the dependence on the neural populations in the drift and diffusion functions. We therefore introduce three maps:
\begin{itemize}
	\item the  measurable functions $f: \Gamma \times \RR \times E\mapsto E$ and $g: \Gamma \times \RR \times E \mapsto E^m$ 
	\item the map $b:\Gamma\times \Gamma \times \RR \times E \times E\mapsto E$ which is assumed measurable,
\end{itemize} 
and rewrite the network equations as:
\begin{multline}
 dX_t^{i,N}=f(r_{\alpha},t,X_t^{i,N})\\+\frac 1 {P(N)} \sum_{\gamma=1}^{P(N)}\sum_{p(j)=\gamma}\frac1{N_\gamma}b(r_{\alpha},r_{\gamma},w_{ij},X_t^{i,N},X_{t-\tau_{ij}}^{j,N})dt+g(r_{\alpha},t,X_t^{i,N})\cdot dW_t^i,
 \label{eq:NetworkSpace}
\end{multline}
These equations are clearly well-defined as proved in proposition~\ref{prop1}. As described in the macroscopic framework~\ref{sec:setting}, the two sequences of random variables $(w_{ij})$ and $(\tau_{ij})$ for fixed $i\in\N$ are independent, and for fixed $(i,j)$, $\tau_{ij}$ and $w_{ij}$ are correlated. Their law depend on the locations $r_{\alpha}$ and $r_{\gamma}$ of the microcolumns neurons $i$ and $j$ belong to. We denote $\Lambda_{r_{\alpha},r_{\gamma}}$ this law. We assume that this law is measurable with respect to the Borel algebra of $\Gamma$, i.e. for any $A \in \B(\R\times \R_+)$ the Borel algebra of $\R\times \R_+$, the map $(r,r')\mapsto \Lambda_{r,r'} (A)$ is measurable with respect to $\B(\Gamma\times \Gamma)$.  We assume that assumptions (H1)-(H4) are valid uniformly in the space variables, and consider the \emph{neural field limit} given by the condition:
\begin{equation}\label{eq:NeuralFieldLimit}
	\varepsilon(N)=\frac 1 {P(N)} \sum_{\gamma=1}^{P(N)} \frac 1 {N_{\gamma}}\mathop{\longrightarrow}\limits_{N\to\infty} 0.
\end{equation}
Elaborating on the proofs provided (i) in the finite-population case treated in the present manuscript and (ii) in the neural field limit for non random synaptic weights or delays, we will show that the network equations converge towards a spatially-extended McKean-Vlasov equation:
\begin{multline}
 dX_t(r)=f(r,t,X_t(r))\,dt+g(r,t,X_t(r))\cdot dW_t(r)\\
+\int_{\Gamma}\int_{\R}\int_{-\tau}^0 \EE_{Z}[b(r,r',j,X_t(r),Z_{t+s}(r'))] d\Lambda_{r,r'}(j,s) d\lambda (r') dt.
 \label{eq:MeanFieldSpace}
\end{multline}
In these equations, the process $(W_t(r))$ is a chaotic Brownian motion (as defined in~\cite{touboulNeuralfields:11}), i.e. a stochastic process indexed by space $r\in\Gamma$, such that for any $r\in\Gamma$, the process $W_t(r)$ is a standard $m$-dimensional Brownian motion and for any $r\neq r' \in \Gamma^2$, $W_t(r)$ and $W_t(r')$ are independent. These processes are singular functions of space, and in particular not measurable with respect to the Borel algebra of $\Gamma$, $\B(\Gamma)$. Therefore, the solutions are themselves not measurable, which raise questions on the definition of the mean-field equation~\eqref{eq:MeanFieldSpace} in particular for the definition of the integral on space of the mean-field term. However, it was shown in~\cite{touboulNeuralfields:11}, making sense of this equation amounts showing that the law of the solution is $\B(\Gamma)$-measurable. Once this is proved, the integral is well defined. In the spatial case, we make the following assumptions, that are directly corresponding to the assumptions (H1)-(H4) of the finite-population case:
 \begin{itemize}
 \item[(H1')] $f$ and $g$ are uniformly Lipschitz-continuous functions with respect to their last variable.
 \item[(H2')] For almost all $w\in\R$ and any $(r,r^\prime)\in\Gamma^2$, $b(r,r',w,\cdot,\cdot)$ is $L$-Lipschitz-continuous, i.e. for any $(x,y)$ and $(x^\prime,y^\prime)$ in $E\times E$, we have: $$|b(r,r',w,x,y)-b(r,r',w,x^\prime,y^\prime)|\leq L(|x-x^\prime|+|y-y^\prime|).$$
 \item[(H3')] There exists a function $\bar K : \R\mapsto \RR^+$ such that for any $(r,r')\in\Gamma^2$,  $$|b(r,r',w,x,y)|^2\leq\bar K(w)\qquad\mbox{and}\qquad\E_{\Lambda_{r,r'}}[\bar K(w)]\leq\bar k<\infty.$$
 \item[(H4')] The drift and diffusion functions satisfy the uniform (in $r$) monotone growth condition: 
$$x^Tf(r,t,x)+\frac12|g(r,t,x)|^2\leq K(1+|x|^2).$$
\end{itemize}
The initial conditions we consider for the mean-field equations are processes $(\zeta_t(r), t\in [-\tau,0])\in \mathcal{X}_0$ the space of spatially chaotic square integrable process with measurable law, processes such that the regularity conditions are satisfied:
\begin{itemize}
	\item for any $r\in\Gamma$, $\zeta_t(r)$ is square integrable in $\mathcal C_{\tau}$ 
	\item for any $r\neq r'$, the processes $\zeta(r)$ and $\zeta(r')$ are independent
	\item for fixed $t\in[-\tau,0]$, the law of $\zeta_t(r)$ is measurable with respect to $\B(\Gamma)$, i.e. for any $A\in \B(E)$, $p_{\zeta_t}(r)=\PP(\zeta_t(r)\in A)$ is a measurable function of $(\Gamma,\B(\Gamma))$ in $[0,1]$. 
\end{itemize}
We will denote $\mathcal{X}_T$ the set of processes $(\zeta_t(r), t\in [-\tau,T])$ satisfying the above regularity conditions on $[-\tau, T]$.

\begin{prop}\label{prop:ExistenceUniquenessSpace}
	Under assumptions (H1')-(H4'), for any initial condition $\zeta\in \mathcal{X}$, there exists a unique, well-defined strong solution to the mean-field equations~\eqref{eq:MeanFieldSpace}.
\end{prop}
	The proof classically starts by showing square integrability of possible solutions, then considers equation~\eqref{eq:MeanFieldSpace} as a fixed point equation $X_t=\Phi(X_t)$, and shows a convergence property of iterates of the map $\Phi$ starting from an arbitrary chaotic process $X^0_t(r)\in \mathcal{X}_T$. It is easy to see that the function $\Phi$ maps $\mathcal{X}_T$ in itself. The sequence of processes $X^k=\Phi^k(X^0)$ is therefore well-defined. Estimates similar to those proved in proposition~\ref{prop1} and theorem~\ref{theo1} allow concluding on the existence and uniqueness of solutions. The proof being classical, it is left to the interested reader extending the argument of~\cite[Theorem 2]{touboulNeuralfields:11} to our random environment setting. 

The convergence result of the network equations towards the mean-field equations can be stated as follows:
\begin{theo}\label{thm:ConvergenceSpace}
	Let $\zeta \in \mathcal{X}_0$ a chaotic process. Consider the process $(X^{i,N}_t, t\in[-\tau,T])$ solution of the network equations~\eqref{eq:NetworkSpace} with independent initial conditions identically distributed for neurons in the same population located at $r\in\Gamma$ with law equal to $(\xi_t(r),t\in[-\tau,0])$. Under assumptions (H1')-(H4') and the neural field limit assumption~\eqref{eq:NeuralFieldLimit}, the process $(X^{i,N}_t, t\in[-\tau,T])$ converges in law towards $(X_t(r), t\in [-\tau,T])$ solution of the mean-field equations with initial conditions $\zeta$.
\end{theo}
The proof of this result proceeds as that of~\cite[Theorem 3]{touboulNeuralfields:11} including the refinements brought in the proof of theorem~\ref{theo1} to take into account random connectivities and delays.


\begin{thebibliography}{10}
\providecommand{\url}[1]{{#1}}
\providecommand{\urlprefix}{URL }
\expandafter\ifx\csname urlstyle\endcsname\relax
  \providecommand{\doi}[1]{DOI~\discretionary{}{}{}#1}\else
  \providecommand{\doi}{DOI~\discretionary{}{}{}\begingroup
  \urlstyle{rm}\Url}\fi

\bibitem{aradi-soltesz:02}
Aradi, I., Soltesz, I.: Modulation of network behaviour by changes in variance
  in interneuronal properties.
\newblock The Journal of physiology \textbf{538}(1), 227 (2002)

\bibitem{babb-pretorius-etal:89}
Babb, T., Pretorius, J., Kupfer, W., Crandall, P.: Glutamate
  decarboxylase-immunoreactive neurons are preserved in human epileptic
  hippocampus.
\newblock Journal of Neuroscience \textbf{9}(7), 2562--2574 (1989)

\bibitem{bassett2006small}
Bassett, D.S., Bullmore, E.: Small-world brain networks.
\newblock The neuroscientist \textbf{12}(6), 512--523 (2006)

\bibitem{bettus-chauvel-etal:08}
Bettus, G., Wendling, F., Guye, M., Valton, L., R{\'e}gis, J., Chauvel, P.,
  Bartolomei, F.: Enhanced eeg functional connectivity in mesial temporal lobe
  epilepsy.
\newblock Epilepsy research \textbf{81}(1), 58--68 (2008)

\bibitem{bosking-zhang-etal:97}
Bosking, W., Zhang, Y., Schofield, B., Fitzpatrick, D.: Orientation selectivity
  and the arrangement of horizontal connections in tree shrew striate cortex.
\newblock The Journal of Neuroscience \textbf{17}(6), 2112--2127 (1997)

\bibitem{bressloff:12}
Bressloff, P.C.: Spatiotemporal dynamics of continuum neural fields.
\newblock Journal of Physics A: Mathematical and Theoretical \textbf{45}(3),
  033,001 (2012)

\bibitem{bullmore2009complex}
Bullmore, E., Sporns, O.: Complex brain networks: graph theoretical analysis of
  structural and functional systems.
\newblock Nature Reviews Neuroscience \textbf{10}(3), 186--198 (2009)

\bibitem{buszaki:06}
Buzsaki, G.: Rhythms of the brain.
\newblock Oxford University Press, USA (2004)

\bibitem{da-prato:92}
Da~Prato, G., Zabczyk, J.: Stochastic equations in infinite dimensions.
\newblock Cambridge Univ Pr (1992)

\bibitem{dobrushin:70}
Dobrushin, R.: Prescribing a system of random variables by conditional
  distributions.
\newblock Theory of Probability and its Applications \textbf{15} (1970)

\bibitem{ermentrout-cowan:79}
Ermentrout, G., Cowan, J.: {Temporal oscillations in neuronal nets}.
\newblock Journal of mathematical biology \textbf{7}(3), 265--280 (1979)

\bibitem{ermentrout-terman:10b}
Ermentrout, G., Terman, D.: Mathematical foundations of neuroscience (2010)

\bibitem{ermentrout-terman:10}
Ermentrout, G.B., Terman, D.: Foundations Of Mathematical Neuroscience.
\newblock Interdisciplinary Applied Mathematics. Springer (2010)

\bibitem{fitzhugh:55}
FitzHugh, R.: Mathematical models of threshold phenomena in the nerve membrane.
\newblock Bulletin of Mathematical Biology \textbf{17}(4), 257--278 
  0092--8240 (1955)

\bibitem{gray1989oscillatory}
Gray, C.M., K{\"o}nig, P., Engel, A.K., Singer, W., et~al.: Oscillatory
  responses in cat visual cortex exhibit inter-columnar synchronization which
  reflects global stimulus properties.
\newblock Nature \textbf{338}(6213), 334--337 (1989)

\bibitem{hodgkin-huxley:39}
Hodgkin, A., Huxley, A.: {Action potentials recorded from inside a nerve
  fibre}.
\newblock Nature \textbf{144}, 710--711 (1939)

\bibitem{hodgkin-huxley:52}
Hodgkin, A., Huxley, A.: A quantitative description of membrane current and its
  application to conduction and excitation in nerve.
\newblock Journal of Physiology \textbf{117}, 500--544 (1952)

\bibitem{jansen-rit:95}
Jansen, B.H., Rit, V.G.: Electroencephalogram and visual evoked potential
  generation in a mathematical model of coupled cortical columns.
\newblock Biological Cybernetics \textbf{73}, 357--366 (1995)

\bibitem{mao:08}
Mao, X.: Stochastic Differential Equations and Applications.
\newblock Horwood publishing (2008)

\bibitem{munoz-mendez-etal:07}
Munoz, A., Mendez, P., DeFelipe, J., Alvarez-Leefmans, F.: Cation-chloride
  cotransporters and gaba-ergic innervation in the human epileptic hippocampus.
\newblock Epilepsia \textbf{48}(4), 663--673 (2007)

\bibitem{noebels:96}
Noebels, J.: Targeting epilepsy genes minireview.
\newblock Neuron \textbf{16}, 241--244 (1996)

\bibitem{ohki-chung-etal:05}
Ohki, K., Chung, S., Ch'ng, Y., Kara, P., Reid, R.: Functional imaging with
  cellular resolution reveals precise micro-architecture in visual cortex.
\newblock Nature \textbf{433}, 597--603 (2005)

\bibitem{schnitzler-gross:05}
Schnitzler, A., Gross, J.: Normal and pathological oscillatory communication in
  the brain.
\newblock Nat. Rev. Neurosci. \textbf{6}(4), 285--296 (2005)

\bibitem{shpiro2007dynamical}
Shpiro, A., Curtu, R., Rinzel, J., Rubin, N.: Dynamical characteristics common
  to neuronal competition models.
\newblock Journal of neurophysiology \textbf{97}(1), 462--473 (2007)

\bibitem{sznitman:84a}
Sznitman, A.: Nonlinear reflecting diffusion process, and the propagation of
  chaos and fluctuations associated.
\newblock Journal of Functional Analysis \textbf{56}(3), 311--336 (1984)

\bibitem{sznitman:89}
Sznitman, A.: Topics in propagation of chaos.
\newblock Ecole d'Et{\'e} de Probabilit{\'e}s de Saint-Flour XIX pp. 165--251
  (1989)

\bibitem{tanaka:78}
Tanaka, H.: Probabilistic treatment of the boltzmann equation of maxwellian
  molecules.
\newblock Probability Theory and Related Fields \textbf{46}(1), 67--105 (1978)

\bibitem{touboulNeuralFieldsDynamics:11}
Touboul, J.: On the dynamics of mean-field equations for stochastic neural
  fields with delays

\bibitem{toubouldelays:12}
Touboul, J.: Limits and dynamics of stochastic neuronal networks with random
  delays.
\newblock Journal of Statistical Physics \textbf{149}(4), 569--597 (2012)

\bibitem{touboulNeuralfields:11}
Touboul, J.: The propagation of chaos in neural fields.
\newblock Annals of Applied Probability (in revision)  (2013)

\bibitem{touboul-hermann-faugeras:11}
Touboul, J., Hermann, G., Faugeras, O.: Noise-induced behaviors in neural mean
  field dynamics.
\newblock SIAM J. on Dynamical Systems \textbf{11}(49--81) (2011)

\bibitem{wilson-cowan:72}
Wilson, H., Cowan, J.: Excitatory and inhibitory interactions in localized
  populations of model neurons.
\newblock Biophys. J. \textbf{12}, 1--24 (1972)

\bibitem{wilson-cowan:73}
Wilson, H., Cowan, J.: A mathematical theory of the functional dynamics of
  cortical and thalamic nervous tissue.
\newblock Biological Cybernetics \textbf{13}(2), 55--80 (1973)

\end{thebibliography}
\end{document}